\documentclass{ieeeaccess}
\usepackage{cite}
\usepackage{amsmath,amssymb,amsfonts}
\usepackage{graphicx}
\usepackage{textcomp}

\usepackage{lineno,hyperref}
\usepackage[T1]{fontenc}

\usepackage{xcolor}
\usepackage{amssymb}
\usepackage{float}

\usepackage[utf8]{inputenc}
\usepackage [english]{babel}
\usepackage [autostyle, english = american]{csquotes}
\MakeOuterQuote{"}

\usepackage{algorithm}

\usepackage[noend]{algorithmic}

\newcommand{\TITLE}[1]{\item[#1]}

\newbox\fixbox
\renewcommand{\algorithmicdo}{\setbox\fixbox\hbox{\ {} }\hskip-\wd\fixbox}
\newcommand{\algcost}[2]{\strut\hfill\makebox[1.5cm][l]{#1}\makebox[4cm][l]{#2}}

\newcommand{\nonl}{\renewcommand{\nl}{\let\nl\oldnl}}

\usepackage{textcomp} 
\usepackage{graphicx}
\usepackage{pdfpages}

\usepackage{amsmath}
\DeclareMathOperator*{\Max}{Max}
\newtheorem{theorem}{Theorem}
\newtheorem{lemma}{Lemma}

\usepackage{mathtools} 

\def\BibTeX{{\rm B\kern-.05em{\sc i\kern-.025em b}\kern-.08em
    T\kern-.1667em\lower.7ex\hbox{E}\kern-.125emX}}
\begin{document}
\history{Date of publication xxxx 00, 0000, date of current version xxxx 00, 0000.}
\doi{xxxx.DOI}


\title{Novel Relay Selection Algorithms for Machine-to-Machine Communications with Static RF Interfaces Setting}
\author{\uppercase{Monireh~Allah~Gholi~Ghasri}\authorrefmark{1},
\uppercase{Ali~Mohammad~Afshin~Hemmatyar\authorrefmark{2}, \uppercase{Siavash~Bayat\authorrefmark{3}, and Mostafa~Mahdieh}.\authorrefmark{4},}}
\address[1]{Department
of Computer Engineering, Sharif University of Technology, Azadi Ave., Tehran, Iran (e-mail: moghasri@ce.sharif.edu)}
\address[2]{Department
of Computer Engineering, Sharif University of Technology, Azadi Ave., Tehran, Iran (e-mail: hemmatyar@sharif.edu)}
\address[3]{Electronics Research Institute, Sharif University of Technology, Azadi Ave., Tehran, Iran (e-mail: bayat@sharif.edu)}
\address[4]{Department
of Computer Engineering, Sharif University of Technology, Azadi Ave., Tehran, Iran (e-mail: mahdieh@sharif.edu)}

\markboth
{M. A. G. Ghasri \headeretal: Novel Relay Selection Algorithms for Machine-to-Machine Communications with Static RF Interfaces Setting}
{M. A. G. Ghasri \headeretal: Novel Relay Selection Algorithms for Machine-to-Machine Communications with Static RF Interfaces Setting}

%

\corresp{Corresponding author: Ali~Mohammad~Afshin~Hemmatyar (e-mail: hemmatyar@sharif.edu).}

\begin{abstract}
Machine-to-Machine (M2M) communications play a significant role in the Internet of Things (IoT). Cooperation of machines in M2M communications can improve network performance when the quality of connections between sources and their destinations is poor. Careful selection of machines as relays can play an effective role in improving the quality of communication in dense networks. Furthermore, the possibility of simultaneous use of different Radio Frequency (RF) interfaces can increase the capacity of data transmission over the network. In this paper, two novel M2M relay selection algorithms are proposed namely Optimal Relay Selection Algorithm (ORSA) and Matching based Relay Selection Algorithm (MRSA). ORSA is a centralized algorithm that transforms the relay selection problem to a k-cardinality assignment problem that can be solved using the Hungarian algorithm. MRSA is a decentralized algorithm that leverages concepts from matching theory to provide a stable relay assignment. In both proposed algorithms, static RF interfaces setting is applied to enable the simultaneous use of different interfaces. It is shown that ORSA results are optimal and MRSA results are stable. The simulations compare the capacity of data transmission of proposed and baseline algorithms. The results show that when the number of channels is not restricted, MRSA is only about 3\% lower than ORSA and its results are higher than direct transmission and random relay selection, about 56\% and 117\%, respectively.
\end{abstract}


\titlepgskip=-15pt

\maketitle

\section{Introduction}\label{sec:introduction}
\PARstart{M}{achine}-to-Machine (M2M) communications is a notable part of the Internet of Things (IoT) \cite{iMDHLaSh2015}. In this kind of communication, machines can communicate with each other, without or with minimal human intervention. Cooperation between machines will improve the data rate and the bandwidth efficiency of wireless networks such as Long-Term Evolution-Advanced (LTE-A) cellular networks \cite{iMDHLaSh2015, M3GLGhHs2015, rCNOAsKh2019}.






In the cooperation of machines, some machines can act as relays between other machines. Especially when the direct link between a source and its destination (e.g. the base station) is weak, relays may save transmission power \cite{RMATZhYu2019} and increase network coverage and performance \cite{RMATZhYu2019, DTRWWang2019, rCNOAsKh2019, ORLAbula2012, CRRMBaLo2011}. 
Appropriate assignment of relays to machines is a challenge for dense network communications, which is referred to as \textit{relay selection problem} in this paper. This assignment of relays depends on various network condition parameters, such as Signal-to-Noise Ratio (SNR) which increases the complexity of this problem.



Some studies have used game theory and matching theory to solve relay selection problems in different networks \cite{TMEDXuFe2017, DSRMLiXu2016, CRRMBaLo2011, GTWNZhGu2011}. Matching theory can provide an appropriate framework for analysis and designing the decentralized methods for interactions between the rational and selfish players \cite{MTAWBaLi2016}. There are some studies that have applied matching theory to model the relay selection problem \cite{IRSTZhXu2013, CRLWZhTa2020}. Some works have modeled the relay selection problem as a maximum weighted matching problem in bipartite weighted graphs and the Hungarian algorithm has been applied in the relay selection process, as a solution \cite{GRSCAlFo2011, OCROLiT2012, IRSTZhXu2013, IHRRKiDo2014,  ICRDchitr2016,  ENMRChHu2018, DSTRKaLu2019, DRDHLiWa2019 }.


In M2M communications, machines can be equipped with different radio frequency (RF) interfaces, such as Long Term Evolution (LTE), Bluetooth, Wireless Fidelity (WiFi), or multiple of them simultaneously. Machines can use their RF interfaces to support the different Quality of Services (QoS) and different requirements of the users \cite{iMDHLaSh2015}. The simultaneous use of different RF interfaces to transmit data from a machine to other machines or base station can increase M2M communications average throghput in a very dense network \cite{PORISiJK2016, AMHCHuXu2017}. The authors are not aware of any relay selection algorithms for M2M communications which considered the possibility of machines using multiple RF interfaces simultaneously. 


This simultaneous RF interfaces setting can be static or dynamic. For example, M2M communications can use WiFi interface to transmit data among machines and they can use LTE interface for machine to base station communications, statically. In another case, machines may use WiFi, Bluetooth or ZigBee interfaces to transmit data among machines, dynamically.

 For instance, a smart city can be considered. In this city, there are several uses for sending data by machines. In some of these applications, fixed machines send their data at different times using RF interfaces such as LTE or WiFi. As an example, smart urban lighting systems can be mentioned as an application in the smart city. These systems are turned on and off only at certain hours of the day and night. Therefore, these systems usually do not require data exchange at other times.

%

Security systems such as anti-theft systems or fire alarms in the city also need to send data to a security center such as the fire department or the police. In this case, the machines are idle most of the time and at certain times need to transmit data quickly to the base stations. At the time of data transmission, due to channel conditions, the direct connection of each source with the base station may not be appropriate. In this case, other devices in the smart city that are not currently sending data can be utilized as relays. In this manner, the sources send their data to the base station through the relays and appropriate relay selection will improve the network transmission quality.


%

\subsection{Related Works}\label{subsec:relWork}

Relay selection may be useful in order to forward data to the base station to deliver to its destination. The network condition parameters, such as SNR or Signal-to-Interference-plus-Noise-Ratio (SINR), can be involved in selecting the appropriate relay \cite{PMESHuWe2017, rCNOAsKh2019}. This relay selection can be very important, especially when the direct link between a source and its destination (e.g. the base station) is weak \cite{CRRMBaLo2011}, the network coverage needs to be extended \cite{sBARNoCh2016, ORLAbula2012}, and the transmission power should be reducesd for increasing life time of the machine \cite{IAEMTsMi2017, IEPCTsPa2017}. Thus use of relays can increase network throughput \cite{sBARNoCh2016}. In this subsection, some of the previous related works are summarized. 




Recently, a relay selection algorithm based on the Basic Sequential Algorithmic Scheme (BSAS) is proposed for high density LTE networks \cite{RSCLHaAl2019}. Two layers of users are considered, in this work. The first layer users are directly connected to the base station and the second layer users use one of the first layer users as a relay to connect the base station. In this algorithm, the users form clusters and each cluster has a cluster head from the first layer nodes. These cluster heads transmit data of all other users in its cluster. The proposed algorithm improves the system capacity and energy consumption compared to other similar work \cite{RSCLHaAl2019}.

Another paper provided two new approaches to modify the buffer-aided relay selection algorithms in equal maximum weight link conditions \cite{barcRaJa2018}. The authors proposed two metrics to use for this condition in each of the new approaches. The first parameter is used in one of the approaches is SNR. The results show that involving this parameter in relay selection improves the outage probability. The other parameter is prioritizing links between relays and destinations based on the occupied buffer space. Involving this parameter in the second approach can improve the delay and throughput performances \cite{barcRaJa2018}.


%
%
%
%




In the other work, two relay selection schemes based on two different parameters, Signal-to-Interference Ratio (SIR) or location,for Machine-Type Communication (MTC) are proposed \cite{rrmcTeLi2017}. In this paper, gateways, as relays, receive MTC devices data and transmit it to the base station. In the relay selection based on SIR, gateways attempt to receive data from MTC transmitters that have the highest received power. The possibility of simultaneous data transmission by multiple MTC transmitters in this schema can lead to high interference and reduces the probability of successful data decoding. Furthermore, the data of each MTC transmitter may be received by the base station through multiple gateways. The relay selection based on location modifies the SIR based scheme, by assigning the nearest MTC transmitter to each gateway and farther MTC transmitters blocked by this gateway. Thus despite the cost of sending spatial data by MTC transmitters, the received interference by each gateway is reduced and the duplicate MTC data transmission to the base station is avoided \cite{rrmcTeLi2017}.

	\subsubsection{Hungarian based relay selection}
	
The Hungarian algorithm is a solution for the maximum weighted matching problem in bipartite weighted graph \cite{ICRDchitr2016}. Following, a review of Hungarian algorithm based relay selection schemes are mentioned:


A relay selection algorithm has been proposed with subchannel reusing in the Device-to-Device (D2D) communications \cite{DSTRKaLu2019}. In this algorithm, a graph coloring algorithm is applied to arrange the D2D peers into nonconflicting groups that have minimum intergroup interference. Then a matrix of D2D peers power consumption is constructed, and relay selection of D2D peers is formulated as a weighted bipartite graph matching problem. In the next phase, this problem is solved by the Hungarian algorithm \cite{DSTRKaLu2019}.

A study has been conducted on the effectiveness of relay selection in 3GPP Narrowband networks in Internet of Things (NB-IoT)  \cite{ENMRChHu2018}. To increase the chance of successful transmission, NB-IoT adopts a repetition-based transmission. To reduce the number of repetitions,  relay selection can be utilized. In this work, relay selection is modeled as a weighted bipartite matching problem and a solution is obtained using the Hungarian algorithm \cite{ENMRChHu2018}.


The assignment problem and relay selection for the relay-aided D2D communications underlying cellular networks has been studied. It is known that this problem is NP-complete, therefore researchers have proposed an Iterative Hungarian Method (IHM) to obtain a near-optimal solution for this problem \cite{IHRRKiDo2014}.

A joint relay selection and resource allocation algorithm is investigated in cognitive networks \cite{GRSCAlFo2011}. In their study, these problems are modeled by bipartite weighted matching in two stages and are solved with the Hungarian algorithm \cite{GRSCAlFo2011}.



			\subsubsection{Matching theory based relay selection}

Matching theory can provide an appropriate framework for analysis and designing the distributed methods for interactions between rational and selfish players \cite{MTAWBaLi2016}. Here, some papers that used from the matching theory in relay selection are summarized.

Jointly optimizing resource management, relay selection, spectrum allocation, and power control is an NP-hard problem \cite{TMEDXuFe2017}. A pricing-based two-stage matching algorithm is provided to maximize energy efficiency. Firstly, a two-dimensional matching is modeled for the spectrum resources reused by relay-to-receiver links. Then, matching users, relays, optimal power control and the spectrum resources reused by transmitter-to-relay links are conducted by a three-dimensional matching \cite{TMEDXuFe2017}.




A distributed satisfaction-aware relay assignment based on the many-to-one matching-game theory is provided \cite{DSRMLiXu2016}. In this work, sources request to relays, with limited resources, regarding their dynamic throughput requirements. Finally, the satisfaction and fairness of sources have been improved \cite{DSRMLiXu2016}.


A distributed matching algorithm to select suitable relays among secondary users for primary users is proposed \cite {CRRMBaLo2011}. The secondary users negotiate with the primary users on the time of both cooperatively relaying the primary users data and allowed spectrum access \cite {CRRMBaLo2011}.

\subsection{Main Contributions}\label{subsec:contribution}

In this paper, two relay selection algorithms are introduced for M2M communications with static RF interfaces setting. In summary, this paper makes the following contributions:
\begin{enumerate}

\item[-] A new solver for the $k$AP is provided by converting it to a standard assignment problem and this new problem is solved by the Hungarian algorithm.  The proof of optimality of this solver is provided in Subsection \ref{sec:k_cardinality_proof}.  

\item[-] A novel centralized Optimal Relay Selection Algorithm (ORSA) is proposed to provide the optimal solution for the relay selection problem in Section \ref{sec:corsa_alg}. The presented method converts the M2M relay selection problem to a $k$-cardinality assignment problem ($k$AP). 


\item[-] A novel decentralized Matching based Relay Selection Algorithm (MRSA) is proposed that is developed by using matching theory. In this algorithm, all nodes (machines and base station) only need local information. The result of this algorithm provides a stable matching between sources, relays, and the base station, based on the \textit{deferred acceptance procedure} \cite{CASMGaSh1962}. The proof of stability of this algorithm is introduced in Appendix \ref{sec:opti_MRSA_proof}. MRSA provides results comparable to the optimal solution and it is applicable in practical situations. 

\item[-] In both of novel relay selection algorithms for M2M communications, the static parallel usage of different RF interfaces is considered. This static RF interfaces setting can enable simultaneous transmission among machines and machines with the base station.
\end{enumerate}

The rest of this article is organized as follows. The system model is described in section \ref{sec:sysModel}. Then, the proposed centralized relay selection algorithm and the proposed decentralized relay selection algorithm are presented in sections \ref{sec:proORSA} and \ref{sec:proMRSA}, respectively. The simulation results are illustrated in section \ref{sec:simulation}. Finally, in section \ref{sec:conclusion}, the conclusion of this paper is provided. 


\section{System Model}\label{sec:sysModel}

We consider a cell with one base station in the center of the cell, and $N$ machines each equipped with two different RF interfaces. In this model, we consider only the uplink paths. The machines have fixed positions, and are divided to two sets, active machines (sources) and idle machines (relays). We denote active machines set as $M^a$ that includes $N^s = | M^a |$ sources and each source wants to send data to the base station to deliver its message to its destination. 

Moreover, the channel between the sources and the base station may have low communication quality, due to path loss, fading and shadowing. We denote the set of idle machines by $M^i$, which contains  $N^r=| M^i |$  relays. In other words, $M_{Total}$ is considered as a set of machines, such that $M_{Total}= M^a  \cup M^i$ and $M^a  \cap M^i= \phi$. According to what was mentioned, $N = N^s + N^r$. It is assumed machines would cooperate with each other. The idle machines do not have data to send at that period of time, thus they can work as a relay. Therefore, when active machines need help, the idle machines can assist them as relays to increase data-rate. Unlike some works where one source can be connected to several relays at the same time \cite{DOCMHuXu2017}, in our model, each source can not be connected to more than one relay.



\begin{figure}[!htb]
\centering
\includegraphics[scale=0.40]{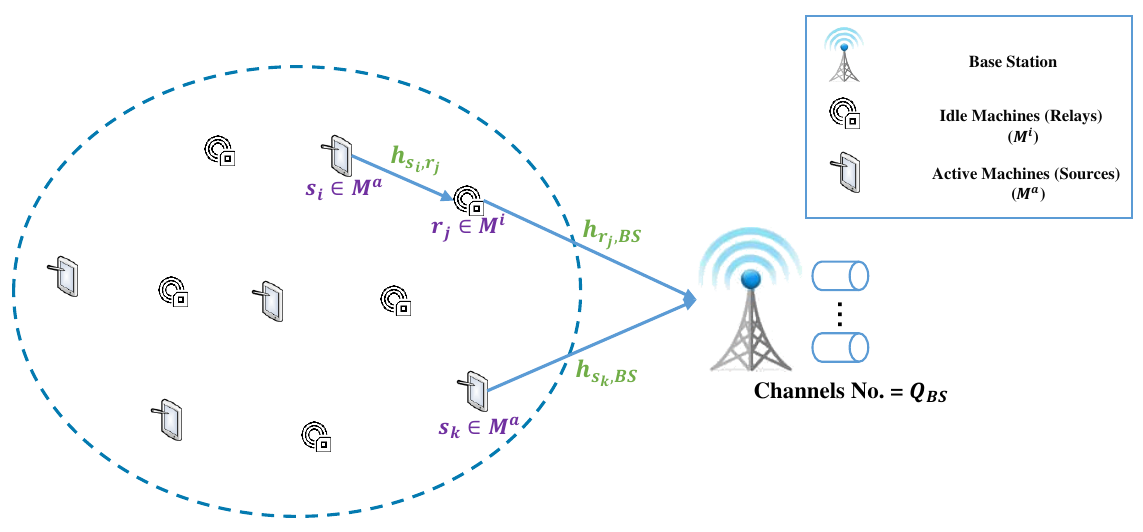}
\caption[The scheme of active and idle machines in the system model.]{The scheme of active and idle machines in the system model.}
\label{fig:plotM2Mpapers}
\end{figure}

 
The relays are using the Decode-and-Forward (DF) protocol. Fig.~\ref{fig:plotM2Mpapers} shows the scheme of active and idle machines in the system model. These machines are randomly positioned with a uniform distribution.



The communication capacity or the maximum bit rate that can be used according to the link conditions between two nodes, $i$ and $j$, in the network is denoted by $C_{i,j}$, that according to the Shannon-Hartley equation, will be as: 

\begin{equation}
\label{eq:Capij}
	C_{i,j} = B^t   log_2(1+\mathrm{SINR}_{i,j}),
\end{equation}	
where $B^t$ is the bandwidth of the communication channel with technology $t$ used by two nodes i and j and $\mathrm{SINR}_{i,j}$ is the SINR of the channel with $i$th node as the transmitter and $j$th node as the receiver.

The possibility of using different RF interfaces of the machines can help to provide high data rates. Therefore, for the static RF interfaces setting, we consider a WiFi interface for their M2M communication and an LTE interface for their direct communication with the base station. The frequency bands used by the LTE and WiFi interfaces are considered different from each other. 

 For the WiFi and LTE technologies, the achieved $\mathrm{SINR}$ of two nodes $i$ and $j$ is computed by the equations (\ref{eq:sinrWifi}) and (\ref{eq:sinrLte}), respectively, which are described as:

\begin{equation}
\label{eq:sinrWifi}
	\mathrm{SINR}_{i,j}^{\mathrm{WiFi}} = \frac{P_i^{\mathrm{WiFi}} \times h_{(i,j)} }{\sigma^2 + \sum_{(k\in M^a, k \neq i)} P_k^{\mathrm{WiFi}} \times h_{(k,j)} },
\end{equation}		
and
\begin{equation}
\label{eq:sinrLte}
	\mathrm{SINR}_{i,j}^{\mathrm{LTE}} = \frac{P_i^{\mathrm{LTE}} \times h_{(i,j)} }{\sigma^2 },
\end{equation}
where $P_i^{\mathrm{WiFi}}$and $P_i^{\mathrm{LTE}}$ are transmission powers of $i$th node for the WiFi and LTE interfaces, respectively, $\sigma^2$ is noise power and $h_{(i,j)}$ is the gain of the channel between the $i$th and $j$th nodes. $h_{(i,j)}$ is modeled by considering path loss, shadowing, and small scale fading. In more exact terms, the following formulation is used:







\begin{align}
\label{eq:channelgain}
    & h_{(i,j)}(dB) 
    \begin{aligned}[t]
       &= Path Loss_{(i,j)}(dB)+ Shadowing (dB) \\
       & + Small  Scale  Fading(dB).  \\
    \end{aligned} \notag \\
\end{align}

where path loss is modeled with path loss exponent $\beta=4$ (equation \ref{eq:pathloss4}). Shadowing (dB)is modeled by a Normal random variable with zero mean and standard deviation of 8 ($\mathcal{N}(0,64)$). Small scale fading (dB) is modeled by a Rayleigh random variable with scale parameter of $\sigma_r = 1$. 


\begin{equation}
\label{eq:pathloss4}
Path Loss_{(i,j)}(dB)=10 \beta log_{10}(\frac{d_{(i,j)}}{d_{0}},
\end{equation}

where $d_{0}=10 (m)$, and $d_{(i,j)}$ is the euclidean distance between node $i$ and node $j$.



The interference experienced at a relay receiver when communicating with a source on the WiFi interface is calculated based on the total power originated from other sources with WiFi interfaces. However, since the WiFi and LTE frequencies are different, the connection between the sources and the relays does not interfere with the connection of the machines (sources or relays) with the base station. Thus the  interference of LTE transmitters at the base station will be zero. It should be mention, to simplify our simulations, we consider maximum probable interference on both RF interfaces, WiFi and LTE. We have simulated the worst possible interference conditions, so it can be expected the results of the real world scenarios would be better than our simulation results.

In this model, any source can select a direct path to the base station or a relay for two hop data forwarding to the base station, by using the matching algorithm. The relays receive the source data on the WiFi interface and send it to the base station on the LTE interface in a time slot. The capacity of the channel between a source and a base station in two hops based on DF relaying is given by \cite{IRSTZhXu2013, RPDUChMa2016}:

\begin{equation}
\label{eq:Capsd}
	C_{i,j} = min \lbrace C_{s,r}, C_{r,d} \rbrace,
\end{equation}
where $C_{s,r}$ is the communication capacity between the source and the relay and $C_{r,d}$ is capacity between the relay and the base station.

\subsection{Problem Formulation}

%
%

In this paper, we propose an algorithm to assign link of the sources to their next hop, which can be either a relay or the base station. 
Some works have considered their selection criteria locally \cite{PMESHuWe2017}, for example based on the choice of the neighbor with the best channel conditions. But we seek to globally maximize the total connection capacity of the sources within the network.
We formulated the problem of link assignment in equation (\ref{eq:optEq_1}). In the following, we presented two solutions for this problem:
\begin{itemize}
\item[-] an optimal relay selection that provides the highest capacity for all network sources,
\item[-] A stable relay selection algorithm that provides a selection which is at least as well as the other possible stable selections for each network source.
\end{itemize}

\begin{align}
\label{eq:optEq_1}
    &\Max_{ x, y, z}
    \begin{aligned}[t]
       &\sum_{i=0}^{N_s-1}{\sum_{j=0}^{N_r-1}\sum_{k=0}^{0}  x_{i,j} y_{j,k} \times min( c_{i,j}, c^{''}_{j,k} )} \\
       & +  		\sum_{i=0}^{N_s-1}\sum_{k=0}^{0} z_{i,k} c^{'}_{i,k},  \\
    \end{aligned} \notag \\\\  
        \text{Subject to} \notag \\
    & x_{i,j} \in \lbrace 0, 1 \rbrace, \quad for \quad 0 \leq i < N_s , \label{eq:optEq_xij} \\   
    & y_{j,k} \in \lbrace 0, 1 \rbrace, \quad for \quad 0 \leq j < N_r, \label{eq:optEq_yij}\\   
    & z_{i,k} \in \lbrace 0, 1 \rbrace, \quad for \quad k=0,  \label{eq:optEq_zij}\\   
    & \sum_{i=0}^{N_s-1} x_{i,j} \leq 1 , \sum_{k=0}^{0} y_{j,k} \leq 1 \quad for \quad 0 \leq j < N_r, \label{eq:optEq_firstnEq}\\
    & \sum_{j=0}^{N_r-1} x_{i,j} \leq 1 , \sum_{k=0}^{0} z_{i,k} \leq 1  \quad for \quad 0 \leq i < N_s , \label{eq:optEq_secnEq}\\    
    &\sum_{i=0}^{N_s-1}{\sum_{j=0}^{N_r-1}\sum_{k=0}^{0}  x_{i,j} y_{j,k}} + \sum_{i=0}^{N_s-1}\sum_{k=0}^{0} z_{i,k} \leq Q_{BS} \label{eq:optEq_forthnEq}.
\end{align}

The definition of the used variables is as follows: 
\begin{itemize}
\item[-] $x_{i,j}$: is $1$ if $i$th source has selected the $j$th relay and $0$ otherwise,
\item[-] $y_{j,k}$: is $1$ if $j$th relay has selected the $k$th base station and $0$ otherwise,
\item[-] $z_{i,k}$ : is $1$ if $i$th source has selected the $k$th base station and $0$ otherwise,
\item[-] $c_{i,j}$: the capacity between $i$th source and $j$th relay,
\item[-] $ c^{'}_{i,k}$: the capacity between $i$th source and $k$th base station,
\item[-] $c^{''}_{j,k}$: the capacity between $j$th relay and $k$th base station, 
\item[-] $N_s$: the number of sources,
\item[-] $N_r$: the number of relays,
\item[-] $Q_{BS}$: quota or connection capacity of the base station. This is equivalent to the number of LTE channels in the simulation scenarios.

%

\item[-] The first summation in inequality (\ref{eq:optEq_firstnEq}) represents the constraint that each relay can only be assigned to a single source.
\item[-] The second summation in inequality (\ref{eq:optEq_firstnEq}) represents the constraint that each relay can only be connected to a single base station. Although this constraint is written in a general form, however we have considered only one base station in our model.



\item[-] The first summation in inequality (\ref{eq:optEq_secnEq}) represents the constraint that each source can only be connected to a single relay.

\item[-] The second summation in inequality (\ref{eq:optEq_secnEq}) represents the constraint that each source can only be connected to a single base station. Although this constraint is written in a general form, however we have considered only one base station in our model.



\item[-] The total number of two hop connections of sources to base station through relays (the first summation in inequality (\ref{eq:optEq_forthnEq})) and total number of direct connections of sources to the base station (the second summation in inequality (\ref{eq:optEq_forthnEq})) is less than or equal to the total number of available channels for connection to the base station ($Q_BS$).


\end{itemize}

\begin{figure}
\centering
\includegraphics[scale=0.45]{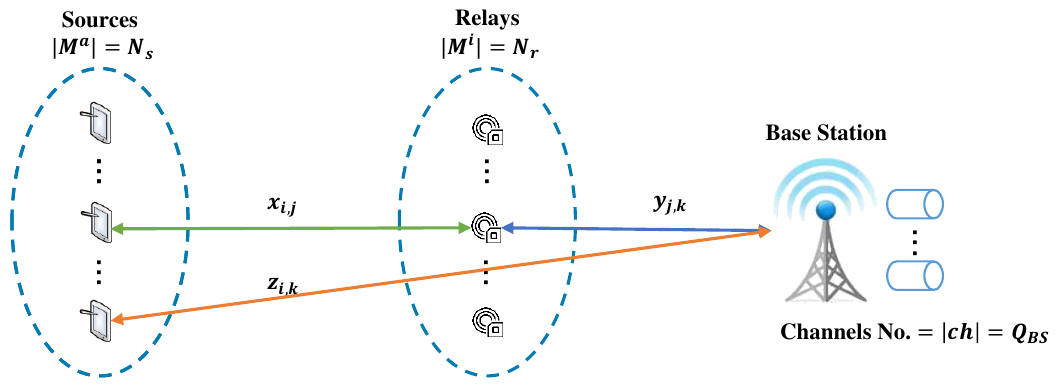}
\caption[The scheme of the graph model of our relay selection problem.]{The scheme of the graph model of our relay selection problem.}
\label{fig:opt_1}
\end{figure}

Fig.~\ref{fig:opt_1} shows the scheme of the graph model of our relay selection problem.

%


\section{Proposed Centralized Relay Selection Algorithm}\label{sec:proORSA}

In this section, we propose a new solver for the $k$-cardinality assignment problem that finds k maximum weighted matching in a bipartite graph. The $k$-cardinality assignment problem can be solved by a polynomial solver \cite{DJThMARC2015}. Some papers proposed solutions \cite{skctVolg2004, kcapDeMa1997}, but we provide a new simple approach for solving it.
In the following, we first describe the proposed solver of the $k$-cardinality assignment problem and then explain the details of the proposed centralized optimal relay selection algorithm. In Subsubsection \ref{sec:k_cardinality_proof}, we prove that the proposed $k$-cardinality assignment solver obtains a mathematically optimal solution for the $k$-cardinality assignment problem.

Then, we use this solver to introduce a centralized relay selection algorithm to solve the problem formulated by equation (\ref{eq:optEq_1}). For this purpose, we transform the relay selection problem to a $k$-cardinality assignment problem and solve the problem using the proposed $k$-cardinality assignment solver. In Section \ref{sec:corsa_alg}, we show that the proposed relay selection algorithm provides an optimal solution for the relay selection problem.

\subsection{A New Solution For The $k$-cardinality Assignment Problem }\label{sec:sol_kCrad}

In this section, we provide a solution for the $k$-cardinality assignment problem, which is a generalization of the standard assignment problem \cite{kcapDeMa1997}. The $k$-cardinality assignment problem is defined as finding the maximum weight matching among all matchings with at most $k$ edges in a bipartite graph. 

The Hungarian algorithm is a common solution for the standard assignment problem \cite{DHACMiSt2007}, but it can not solve a $k$-cardinality assignment problem. Therefore, we transform the $k$-cardinality assignment problem to a standard assignment problem that would be solved by the Hungarian algorithm \cite{KuhaHMAP1955}. The complexity of the algorithm is $O(N^3)$ where $N$ is the number of vertices of the standard assignment problem \cite{DHACMiSt2007, GithFCIH2019}. 

\begin{figure}[htb]
\centering
\includegraphics[scale=0.7]{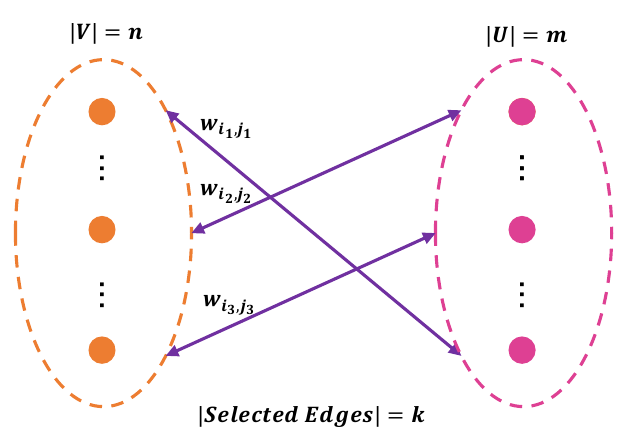}
\caption[The scheme of the bipartite graph model of the $k$-cardinality assignment problem.]{The scheme of the bipartite graph model of the $k$-cardinality assignment problem.}
\label{fig:opt_k_1}

\end{figure}

Our model is a bipartite weighted graph $G=(V \cup U, E)$ where $ \lbrace V \cup U \rbrace $ is set of vertices, $|V|=n$, $|U|=m$, $E=\lbrace (v_i, u_j)| v_i \in V  \wedge  u_j \in U \rbrace $ is the set of edges and  $w_(i, j)$ is the cost of edge $(v_i, u_j)$. We want to select $k$ edges $(k \leq min\lbrace m, n \rbrace )$ so that sum of the weights of selected edges is maximized. If $k=n$ or $k=m$, the new problem will be equivalent to the standard assignment problem without any constraint on the number of selected edges \cite{skctVolg2004, kcapDeMa1997}. The scheme of the bipartite graph model of the $k$-cardinality assignment problem is shown in Fig.~\ref{fig:opt_k_1}.

%


\subsubsection{\textbf{Step 1: Transforming the $k$-cardinality assignment problem to a standard assignment problem}}\label{sec:KtoDef_G}


In the first step, we want to transform the assignment problem with the constraint on the number of edges to a standard assignment problem. In the standard assignment problem, we are looking for a set of edges in the bipartite weighted graph with the maximum total weight. Now, we are going to transform the restricted problem into an unconstrained one, so that the results coming from both problems would be corresponding to each other.

For this purpose, we add some additional vertices to each side. The number of vertices that are added to each side is equal to the difference between the size of another side, $m$ or $n$, and constrained number of edges, $k$. In other words, we add $n_A^{V} = m-k$ vertices to $V$ side and $n_A^{U} = n-k$ vertices to $U$ side.
 

The weight of edges connected to new vertices with other new vertices on the other side is set to $0$ and the weight of other added edges is set to a large enough value ($A_{value}$) such as $1 + \sum_{e \in E} w_e$ or infinity. Intuitively, adding new vertices and their edges by this method causes the lower weight initial edges to be defeated by $A_{value}$-weighted edges. In Lemma \ref{le:nAEq} proved that exactly $((m-k)+(n-k))$ edges with $A_{value}$-weight are selected, therefore, only $k$ initial edges with maximum total weight can be selected in the optimal assignment.

 

%
 
\begin{figure}
\centering
\includegraphics[scale=0.50]{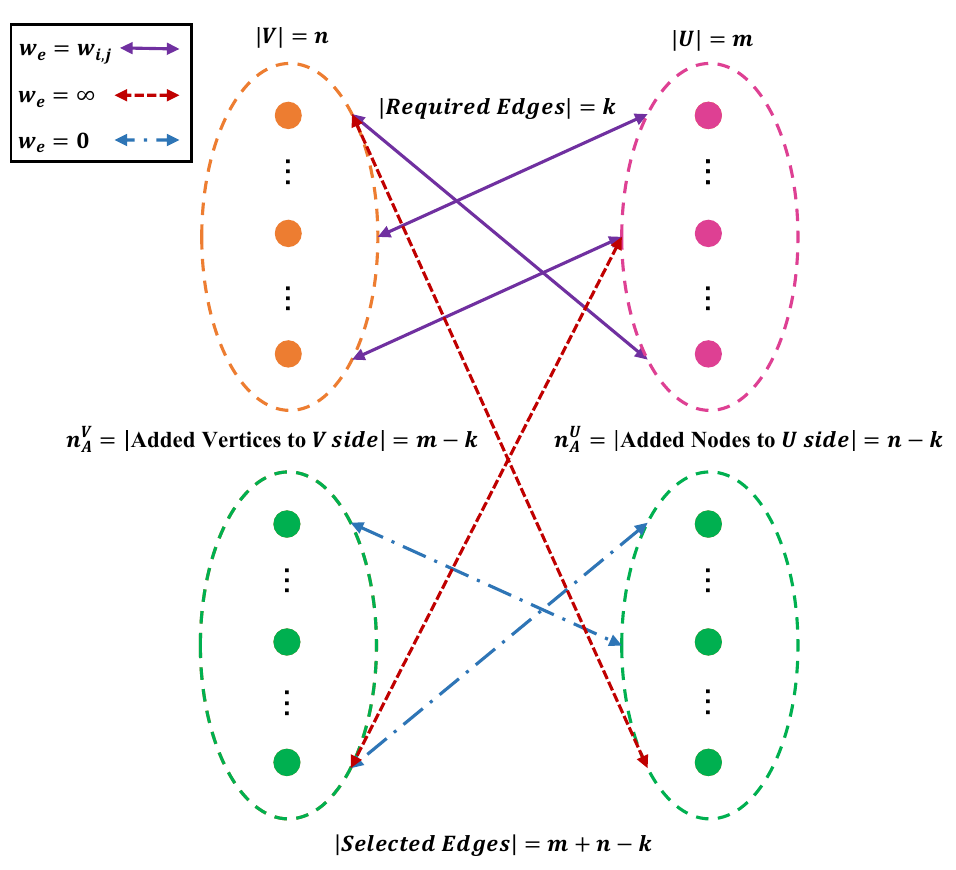}
\caption[The transformation of a $k$-cardinality assignment problem to a standard assignment problem.]{The transformation of a $k$-cardinality assignment problem to a standard assignment problem.}
\label{fig:opt_k_2}

\end{figure}

Now, we can find maximum weighted matching in the new bipartite graph by the Hungarian algorithm. Fig.~\ref{fig:opt_k_2} shows the transformation of a $k$-cardinality assignment problem to a standard assignment problem. 


%

\subsubsection{\textbf{Step 2: Obtaining final results of the transformed assignment problem}}\label{sec:ObtRes_G}

The output of the Hungarian algorithm is a vector with $m+n-k$ size. Every component of this vector represents a vertex of the $V$ side. The value of any component contains the index of a vertex from the $U$ side. Therefore, it is enough to separate the $n$ first components of the output vector that represent the initial vertices and check their content. If the value of the component is less than $m$, it indicates an edge between a vertex of the $V$ side that is represented by the component index and a vertex of the $U$ side that is represented by the component value. Otherwise, if the value of the component is greater than or equal to $m$, it means that this vertex does not have any edge in the original problem. In this way, we can obtain the solution of the $k$-cardinality assignment problem from the solution of the transformed standard assignment problem. The proof of  correctness of the new method for solving the $k$-cardinality assignment problem will be shown in the Theorem \ref{th:AnsCorrespond}.



\subsubsection{\textbf{Proof of optimality of the proposed $k$-cardinality assignment solver}}\label{sec:k_cardinality_proof}
The mathematical formulation of finding a maximum matching with $k$-cardinality ($k$ edges) in a weighted bipartite graph, is expressed in Equation (\ref{eq:maxE}) as:
\begin{align}
\label{eq:maxE}
    &\Max_{| S_{se} |=k}
    \begin{aligned}[t]
       &\sum_{e_{i,j} \in  S_{se}}{ w_{e_{i, j}} }  , \\
    \end{aligned}
\end{align}
where $S_{se}$ is the set of selected edges of the original problem.

The purpose of the discussion in this subsection is to prove that the method presented in Section \ref{sec:sol_kCrad} provides an optimal solution for equation (\ref{eq:maxE}). 

The transformation from equation (\ref{eq:maxE}) to equation (\ref{eq:nA_maxE}) can be summarized as adding new vertices with $0$-weighted and $A_{value}$-weighted edges to each side. It is proved that the optimal solutions of the transformed equation (\ref{eq:nA_maxE}) are in one-to-one correspondence with the optimal solutions of the original equation (\ref{eq:maxE}). The transformed equation is expressed as:
\begin{align}
\label{eq:nA_maxE}
    &\Max_{| S_{se} |= m+n-k}
    \begin{aligned}[t]
      & \lbrace n_{A_{s}} \times A_{value} + \sum_{{e_{i,j}} \in S_{se}}  w_{e_{i, j}} \rbrace ,  \\
    \end{aligned}
\end{align}
where $n_{A_{s}}$ is the number of selected $A_{value}$-weighted edges and  $S_{se}$ is the set of selected edges of the transformed problem.


\begin{lemma} \label{le:nAEq}
For maximization of the equation (\ref{eq:nA_maxE}), the number of selected $A_{value}$-weighted edges, or $n_{A_{s}}$, is a constant value and $n_{A_{s}} = (m-k)+(n-k)$ .
\end{lemma}

\textbf{Proof}: It is demonstrated by contradiction. 

\textbf{Step 1}- We assume $n_{A_{s}} < (m-k)+(n-k)$ and the summation of edge weights is equal to $w_{sum}$. Therefore, at least one of the new vertices does not connect to the initial vertices with a $A_{value}$-weighted edge. Accordingly, for maximization of  equation (\ref{eq:nA_maxE}), we can replace at least an $A_{value}$-weighted edge with the lowest weight edge ($w_{min}$). Hence, the new summation of edge weights $(w_{sum_{new}})$ is greater than or equal to $w_{sum} - w_{min} + A_{value}$. Since $A_{value} > w_{min}$, thus  $w_{sum_{new}} > w_{sum}$. Now, we reach a solution with $n_{A_{s}} \geq (m-k)+(n-k)$. This number of  $A_{value}$-weighted edges contradicts the initial assumption. Therefore, $n_{A_{s}} \nless (m-k)+(n-k)$.

\textbf{Step 2}- Now, we assume $n_{A_{s}} > (m-k)+(n-k)$. But, it is impossible, because, we have only $(m-k)+(n-k)$ new vertices in total. Hence, we can have up to $(m-k)+(n-k)$ of $A_{value}$-weighted edges. It contradicts the previous assumption. Therefore,  $n_{A_{s}} \ngtr (m-k)+(n-k)$.

Hence, it can be concluded that the number of $A_{value}$-weighted edges is equal to $n_{A_{s}} = (m-k)+(n-k)$. $\blacksquare$

Now, we rewrite the equation (\ref{eq:nA_maxE}) as below: 

\begin{align}
\label{eq:Constant_nA_maxE}
    &\Max_{| \lbrace S_{se}   -   E_{A} \rbrace |= k}
    \begin{aligned}[t]
      &  \lbrace \sum_{{e_{i,j}} \in \lbrace S_{se} - E_{A} \rbrace }  w_{e_{i, j}} +  n_{A_{s}}^{*} \times A_{value} \rbrace , \\
    \end{aligned}  
\end{align}
where $n_{A_{s}}^{*}= (n-k)+(m-k)$ is the number of selected $A_{value}$-weighted edges for optimization of the equation (\ref{eq:nA_maxE}), $S_{se}$ is the set of selected edges of the transformed problem and $E_{A}$ is the set of $A_{value}$-weighted edges.

\begin{theorem} \label{th:AnsCorrespond}
Each optimal solution for the $k$-cardinality assignment problem corresponds to an optimal solution for the transformed standard assignment problem and vice versa.
\end{theorem}

\begin{figure}[H]
\centering
\includegraphics[scale=0.6]{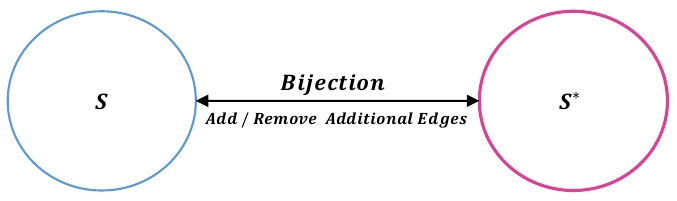}
\caption[The scheme of the bijection between answer space of the problems (\ref{eq:maxE}) and (\ref{eq:Constant_nA_maxE}).]{The scheme of the bijection between answer space of the problems (\ref{eq:maxE}) and (\ref{eq:Constant_nA_maxE}).}
\label{fig:opt_Ans_Space}

\end{figure}

\textbf{Proof}: We denote the set of optimal solutions of equation (\ref{eq:maxE}) as $S$ and the set of optimal solutions of equation (\ref{eq:Constant_nA_maxE}) as $S^*$. Each solution has up to $m+n-k$ edges. According to Lemma \ref{le:nAEq}, the number of $A_{value}$-weighted edges is equal to $n_{A_{s}}^{*} = (n-k)+(m-k)$ in optimal solutions. Hence, the number of edges other than $A_{value}$-weighted edges is equal to $k$. Therefore,

%
%

\begin{itemize}

\item[-] To construct the corresponding item of $S$  from an item of $S^*$, it is enough to remove all the $n_{A_{s}}^{*}$ edges with $A_{value}$-weight and the $k$ remaining obtained edges are the $S$ solution,

and

\item[-] To construct the corresponding item of $S^*$ from an item of $S$, it is enough that $n_ {A_{S}}^{*}$  edges with $A_{value}$-weight from the unassigned initial vertices are connected to the new vertices. Hence, the obtained edges are the $S^*$ solution, and the number of them is equal to $m+n-k$.   $\blacksquare$

\end{itemize}
Fig.~\ref{fig:opt_Ans_Space} shows the scheme of the bijection between the answer space of the problems.

%

%
%

 \subsection{Centralized Optimal Relay Selection Algorithm (ORSA)}\label{sec:corsa_alg}

In this section, an optimal relay selection problem is solved. Thus we transform this problem in two steps to become a $k$AP. In the following, the optimal relay selection problem-solving process is described.

%
%
%

The configuration of our relay selection problem is shown in Fig.~\ref{fig:opt_1}. As it can be seen in this figure, we have two sets of machines, the sources and the relays. The sources should be connected to the base station directly or by using a relay. 



In the assignment, between the sources and relays through the original problem, each source can connect to only one relay and each relay can connect to only one source. Besides, the base station has $Q_{BS}$ channels for communication with the machines, whether a source or a relay. The conditions of the channels are considered similar. On the other words, each source or relay can connect to only one of the channels of the base station, and the base station can connect to up to $Q_{BS}$ machines. Therefore, in the final assignment, the total number of edges connected to the base station can be equal to $Q_{BS}$.

\subsubsection{\textbf{Step 1: Transform our optimal relay selection problem to a $k$-cardinality assignment problem}}\label{sec:RSAtoKcard}

\begin{figure}[!htb]
\centering
\includegraphics[scale=0.7]{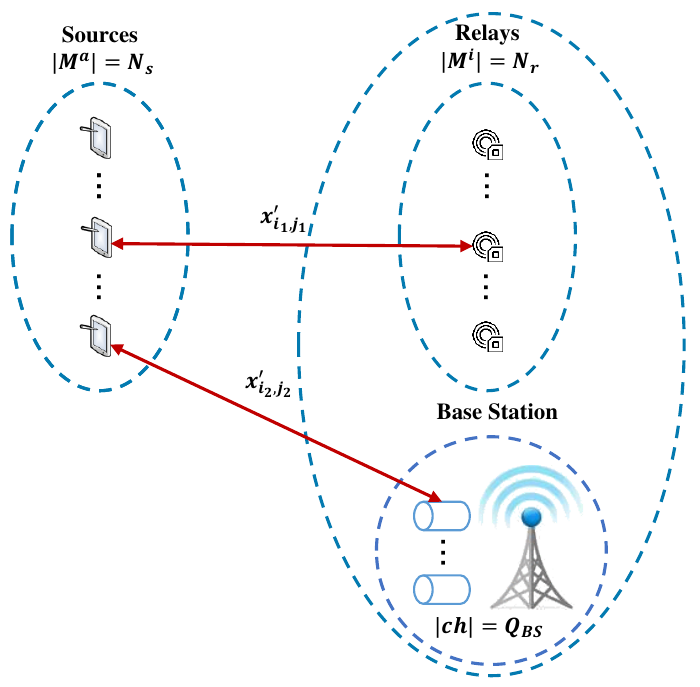}
\caption[The bipartite graph model of the relay selection problem.]{The bipartite graph model of the relay selection problem.}
\label{fig:opt_2}
\end{figure}

In this step, the transformations of the relay selection problem (Fig.~\ref{fig:opt_1} ) to a bipartite graph model (Fig.~\ref{fig:opt_2} ) is presented. To model the relay selection problem as a bipartite graph, we consider $N_s$ vertices on the left side of the bipartite graph, where each vertex corresponds to a source of the network. On the right side of the bipartite graph, we consider $N_r + Q_{BS}$ vertices where the first $N_r$ vertices each correspond to one of the relays and the second $Q_{BS}$ vertices each correspond to one of the $Q_{BS}$ channels of the base station.


The bipartite graph model of the relay selection problem is shown in Fig.~\ref{fig:opt_2}. In this modeling, an edge from the $i$th vertex of the left side to one the top $N_r$ vertices of the right side, represents a connection from the $i_1$th source to a relay, and a connection from that relay to the base station. Furthermore an edge from the $i_2$th vertex of the left side to one of the $Q_{BS}$ vertices represents a direct connection from the $i_2$th source to the base station without any relay. 





The weight of the edge between two vertices on both sides of the graph is defined as follows: 

\begin{itemize}
\item[-] the weight of the edge between a source and a relay is equal to the capacity of two hops path, that is minimum of the capacity of the source and the relay link and the capacity of the relay and the base station link,
and 
\item[-] the weight of the edge between a source and each channel of the base station is equal to the capacity of the source and the base station link.
\end{itemize}

As mentioned, the original problem had the constraint of having at most $Q_{BS}$ connections from the machines to the base station. This constraint is modeled by the selection of at most $Q_{BS}$ edges in the bipartite graph assignment problem. From the modeling presented here, it is clear that the relay selection problem is equivalent to solving the $k$-cardinality assignment problem on this bipartite graph, with $k = Q_{BS}$.

\subsubsection{\textbf{Step 2: Transform our $k$-cardinality assignment problem to a standard assignment problem without edge number constraint}}

We want to find $Q_{BS}$ edges that maximize their total weight in the new problem. The current problem is similar to the problem mentioned in Section \ref{sec:KtoDef_G}. Therefore, it can be solved in the same way. 

\begin{figure}[!htb]
\centering
\includegraphics[scale=0.6]{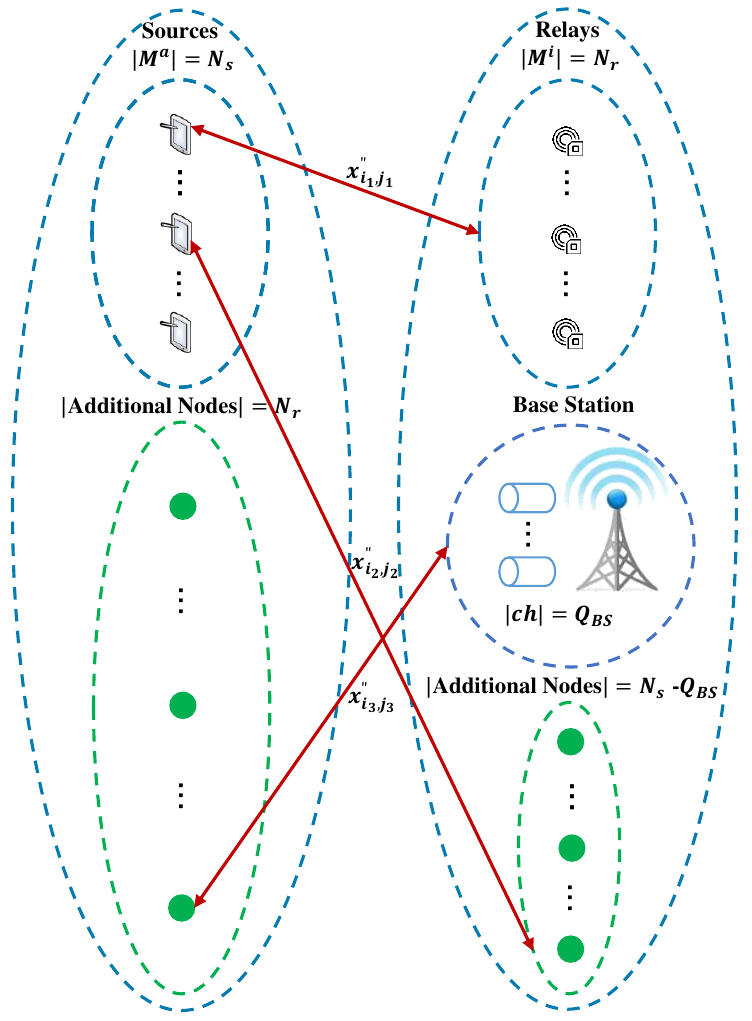}
\caption[The transformed bipartite graph model of the relay selection problem with the additional nodes..]{The transformed bipartite graph model of the relay selection problem with the additional nodes.}
\label{fig:opt_3}

\end{figure}

As stated in Section \ref{sec:KtoDef_G}, we add new vertices to both sides, $N_r + Q_{BS} - Q_{BS}$ new vertices to the left side and $N_s - Q_{BS}$ new vertices to the right side. Now, $N_s + N_r$ vertices exist in both sides, and the new problem is a standard assignment problem that can be solved by a common solution such as the Hungarian algorithm. Fig.~\ref{fig:opt_3} shows the transformed bipartite graph model of the relay selection problem with the additional nodes.


%

\subsubsection{\textbf{Step 3: Obtaining final relay selection from solved assignment problem}}\label{ORSA_ResObtain}

After the two-step transformation, it is necessary to derive the solution of the initial problem from the obtained solution. Similar to Section \ref{sec:ObtRes_G}, we have to extract the corresponding edges of the main problem from the set of output edges of the Hungarian algorithm.
When the Hungarian algorithm is applied to the transformed problem, if it has a solution, its result will be a vector with $N_s + N_r$ components. The content of the $i$th component represents the index of the vertex of the right side ($j$), that is assigned to $i$th vertex of the left side by the Hungarian algorithm. But as mentioned earlier, in Section \ref{sec:ObtRes_G}, the first $N_s$ elements of the result vector are related to the main problem. 


Now, there are three following possible situations for the value of $j$: 

\begin{enumerate}
\item $0 \le 0 j < N_r$,
\item  $N_r \leq j < N_r + Q_{BS}$,
\item  $j \geq N_r + Q_{BS}$.
\end{enumerate}

The first situation indicates that the $i$th source connects to the base station with two hops by the $j$th relay. The second situation indicates that the $i$th source connects to the base station directly. Finally, the last situation indicates that the $i$th source can not connect to any next hop of the network.

\subsubsection{\textbf{Centralized Algorithm Implementation}}

%

This section describes how to implement the transformation of our optimal relay selection problem to a standard assignment problem. We construct a capacity matrix in two steps. To achieve this goal, in the first step, the first part of the matrix is filled by the link capacity of the sources and the relays and each of the base station channels. Hence, we have a capacity matrix of the new $k$AP denoted by $M$. Then, to transform our $k$AP to a standard assignment problem, the rest of the matrix cells are filled by $A_{value}$, that 


\begin{align}
\label{eq:maxE2}
    \begin{aligned}[t]
       &A_{value} >  \sum_{i = 0}^{N_s} \sum_{j = 0}^{N_r+Q_{BS} - 1} M_{i,j} . \\
    \end{aligned}
\end{align}

 For example, $A_{value}$ can be equal to $(max( M_{i, j}) + 1) \times ( N_s+ N_r+Q_{BS})$. 

 
 Now, to find the optimal assignment, the provided capacity matrix is given to an assignment problem solver such as the Hungarian algorithm.  The Hungarian algorithm can be implemented in two different versions, one of which is used by default to find the maximum weighted matching and the other to find the minimum weighted matching. Therefore, if the Hungarian algorithm is implemented to find the maximum weighted matching, the capacity matrix and the number of vertices on each side are given as its inputs. Otherwise, If the Hungarian algorithm is implemented to find the minimum weighted matching, the negative matrix capacity and the number of vertices on each side are given as its inputs.


Finally, the desired output can be obtained from the output of the Hungarian algorithm using the method described in Section \ref{ORSA_ResObtain}. The pseudo code of our proposed centralized matching relay selection algorithm is presented in Algorithm \ref{alg:CMRSA}. 

\subsubsection{\textbf{Complexity of ORSA}}\label{sec:Comp_orsa}
%
%

ORSA is a centralized algorithm. So a central node should apply the proposed centralized relay selection algorithm. In this algorithm, we transform the relay selection problem to a standard assignment problem in two steps. Complexity of the transformation steps (Step 1 and item 1 to 4 of Step 2) is $O((N_s + N_r)^2)$. But complexity of solving the standard assignment problem using the Hungarian algorithm is $O(|vertices||E|)=O((N_s + N_r) \times (N_s + N_r)^2)$ \cite{DHACMiSt2007, GithFCIH2019}. Also Step 3 of Algorithm \ref{alg:CMRSA} has complexity of $O(N_s +  N_r)$. Therefore, the total complexity of ORSA is $O((N_s + N_r)^3)$. If the number of relays ($N_r$) is constant and the number of sources is equal to $N_s = n$, the complexity of ORSA is $O(n^3)$.



\begin{algorithm}[!htb]
\caption{Proposed Centralized Optimum Relay Selection Algorithm.}
\label{alg:CMRSA}
\begin{algorithmic}[1]
\algsetup{linenosize=\small}
\scriptsize
  
%

%

  \TITLE{ \textbf{Step 1: Transform our optimal relay selection problem to a $k$AP}} 

	\STATE Construct the first part of the input capacity matrix of the standard assignment problem, $M_{i, j}$, according to the following rules:
	
	 \begin{itemize}	 
		\item[-] $M_{i, j} = min (C_{s,r}, C_{r, BS}) \quad$	for $(0 \leq i < N_s)$ and $(0 \leq j < N_r)$,  
		\item[-] $M_{i, j} = C_{s,BS}\quad$	for $(0 \leq i < N_s)$ and $(N_r \leq j < N_r+ Q_{BS})$,
	\end{itemize}

  \TITLE{ \textbf{Step 2: Transform our $k$AP to a standard assignment problem and solve it}} 
    
  \STATE $A_{value} = (max( M_{i, j}) + 1) \times ( N_s+ N_r+Q_{BS}) \quad$ for  $(0 \leq i < N_s)$ and $(0 \leq j < N_r+ Q_{BS})$ \algcost{}{}
  	\STATE Construct the second part of the input capacity matrix of the standard assignment problem, $M_{i, j}$, according to the following rules:
	
	 \begin{itemize}	 
		\item[-] $M_{i, j} = A_{value}\quad$	for $(0 \leq i < N_s)$ and $(N_r+ Q_{BS} \leq j < N_s + N_r)$,
		\item[-] $M_{i, j} =  A_{value}\quad$	for $(N_s \leq i < N_s + N_r)$ and $(0 \leq j < N_s + N_r)$.	 
	\end{itemize}

	\STATE Construct the set of edges $E$ of the bipartite graph by $edge_{i, j}$ = (left node index = $i$, right node index = $j$, $M_{i, j}$), 
	\STATE $H^o$ vector = Hungarian(number of vertices $ = N_s+N_r$, edges = $E$ ),  
			
  \TITLE{ \textbf{Step 3:Obtaining final relay selection from solved assignment problem}} 
  
  	\STATE Construct the final output assignment vector, $O$, from the output vector of the standard assignment solution, $H^o$, according to the following rules:
	
	\FOR{$k \gets 1$ to $N$}
		\IF{$H^o_k < N_r$} 
			 \item[-] $O_k = H^o_k$: Meaning that the $k$th source is connected to the base station by the relay with index equal to $O_k$,
		\ELSIF{$ N_r \leq H^o_k < N_r + Q_{BS}$}
			 \item[-] $O_k = N_r$: Meaning that the $k$th source is assigned to the base station directly,
		\ELSIF{$ H^o_k \geq N_r + Q_{BS}$ }
			 \item[-] $O_k = \phi$: Meaning that the $k$th source can not connected to the base station.
		\ENDIF
	\ENDFOR

\end{algorithmic}
\end{algorithm}


\section{Proposed Decentralized Relay Selection Algorithm}\label{sec:proMRSA}

In this section, we propose a decentralized algorithm for relay selection based on matching theory to select a stable selection. First, we describe the matching theory elements, the proposed algorithm players and their preference lists. Then, the matching algorithm is presented to find a stable solution for the equation (\ref{eq:optEq_1}). The proof of stability of matching of MRSA is presented in Appendix \ref{sec:opti_MRSA_proof}.

\subsection{Matching Theory}

Matching theory is a framework to model interaction between the rational and selfish players. We are mapping our problem to a matching theory problem. Some elements of our matching problem are mentioned below.

\subsubsection{The Players}

In our proposed algorithm, there are two types of players defined as follows: 
	\begin{itemize}
		\item Machines consist of sources and relays,
		\item Base station.  
	\end{itemize}

In other words, each machine or the base station in matching algorithm are the rational and selfish players that they want to maximize their communication capacity.

\subsubsection{Utility Function}

The Utility function of machines and the base station in our algorithm is based on the capacity of the paths between the machines and with the base station. As previously mentioned, the capacity of direct and two hop paths are formulated by the equation (\ref{eq:Capij}) and the equation (\ref{eq:Capsd}), respectively. 

\subsubsection{Preference Lists}

The preference lists of machines or base stations are formed according to the node utility functions computed by received channel information between the node and its neighbors. 

\begin{itemize}
\item Each source has a preference list of its neighbor relays and base stations as their candidate next hops.
\item Each relay has two preference lists of its neighbors. The first list, or list of the candidate next hops list, ranks its neighbor base station as next hop for data forwarding of applicant sources. The second list, or the list of the previous hops, rank applicant sources that requested to this relay. For simplicity, in the proposed algorithm, the size of the second list is considered to be one. 
\item Each base station has a list of its neighbor applicant sources and relays that requested it.

We show the preference list of their candidate next hops of the machines by $PL_{NH}$ and the preference list of the applicants of the base station and the relays by $PL_{AP}$.

\end{itemize}

 Any source or relay sort candidate neighbors in the next hop preference list according to capacity in the path consist of this hop. Moreover, any relay or the base station sort the list of applications according to the capacity of the path traversed from that node.

In the following, we study the proposed decentralized relay selection algorithm players and the preference lists.


\subsection{Decentralized Matching based Relay Selection Algorithm (MRSA)}

Due to high density of M2M communications, each machine can have local information from its neighbors and network conditions. On the other hand, although ORSA achieves optimal results, in practice, it can create a bottleneck in dense M2M communications. This bottleneck is either due to communication overhead between the central processor unit and the other nodes or the processing load on the central unit. In this regard, a decentralized algorithm for relay selection may be more suitable for this type of communication. The main idea of the proposed decentralized algorithm for each source is finding a stable matching to select a suitable path, direct path or two hop path, to the base station to reach its data to the destination. To achieve this aim, this algorithm provides a stable best relay selection. 


The weak channel between sources and the base station causes a low data rate in the direct path of sources and the base station. Hence, if a relay is selected as the next hop of a source, the selected neighbor relay has two features. First, it must have enough connection capacity and second, the path containing that relay must have a higher data rate than the rate of the direct path between the source and the base station. Furthermore, the selected relay has an equal or higher data rate than the rate of the paths containing other neighbor relays with enough capacity. 







The proposed decentralized matching based relay selection algorithm (MRSA) is executed in three steps. For simplicity, we assume the machines can synchronize with each other. Some methods are available to synchronize devices in a decentralized network \cite{sSMMBoNy2015, ts5wMaAs2019} which can be used for this manner.


Algorithm \ref{alg:DMRSA1} presents the pseudo code of our proposed decentralized MRSA. At the beginning of the algorithm, the relays broadcast estimated capacity of data transmission with the base station, according to the estimated SINR with it. Therefore, the sources have enough information to sort the candidate next hops, the relays or the base station.

Furthermore, in the initialization section of the algorithm, the machines construct the preference list of their candidate next hops ($PL_{NH}$). In addition, the base station and the relays construct the preference list of their applicants ($PL_{AP}$). Then all of the sources are added to \textit{\textrm{MATCHLIST}}. Any unmatched machines in \textit{\textrm{MATCHLIST}} sort their candidate next hops List, according to channel conditions with their neighbors and received information from them, in the beginning of the matching time.


In step 1, every time the first unmatched machine in \textit{\textrm{MATCHLIST}} requests its application to the first next candidate in its preference list.

On the other hand, each node receiving the request, either a relay or the base station, has a given quota as the connection capacity. The quota for relays is equal to one and the quota of the base station is equal to the number of LTE channels. Thus, any receiver node accepts to a maximum of its quota from the best applicant machines and rejects other machines.

If the relay or base station that received the machine application has enough quota (connection capacity), the applicant machine will be added to the relay or base station preference list. Otherwise, if the new applicant machine is preferred over the worst current applicant machine it will be replaced with it. If neither of these cases hold, the new applicant machine is rejected by the current relay or the base station. Then, the applicant machine requests to its next best  hop candidate, until no candidate remains in its preference list of their candidate next hops.


If a relay that receives a request from a source, does not have a specified next hop, it will be added to \textrm{MATCHLIST} to specify its next hop. Besides, if the relay does not find any next hop, it will reject its applicant sources.



The algorithm continues until no machine remains in \textit{\textrm{MATCHLIST}}. During the execution of step 1 of the algorithm, no machine forwards its data to the destination. In step 2, any nodes match with the final best candidate in preference lists of next hops or applicants. Then, any machine sends data to the base station by the matched next hop that is selected in the previous step. 



%
%


\begin{algorithm}[!htb]
\caption{Proposed Decentralized Matching Based Relay Selection Algorithm: Initialization and Step 1.}
\label{alg:DMRSA1}
\begin{algorithmic}[1]
\algsetup{linenosize=\small}
\scriptsize
  \TITLE{ \textbf{Step 0: Initialization}} 

 \begin{itemize}
		\item[-] Set  $BS= BS_0$,  $M^a = \lbrace All  \quad of  \quad the \quad sources \rbrace$ ,  $M^i = \lbrace All \quad of \quad the \quad relays \rbrace$, and every $m \in \lbrace M^a \cup M^i \rbrace$ is a machine
		\item[-] Set $\textit{\textrm{MATCHLIST}}$ $\leftarrow M^a$, 
		\item[-] Construct \textit{Preference List} of Next Hops ($PL_{NH}$) for the Source,  $PL_{NH} \leftarrow \lbrace  BS_0  \cup \lbrace The \quad neighbor \quad relays \rbrace \rbrace$ ,
		\item[-] Construct \textit{Preference List} of Next Hops ($PL_{NH}$) for the Relays, $PL_{NH} \leftarrow  \lbrace  BS_0  \rbrace$ ,
		\item[-] Construct \textit{Preference  List} of Applicants ($PL_{AP}$) for the Base Stations and the Relays, $PL_{AP} \leftarrow \phi$ ,
 \end{itemize}

  \TITLE{ \textbf{Step 1: Find a suitable next hop for each source}} 

  \WHILE{$\textit{\textrm{MATCHLIST}} \neq \phi$
    \algcost{}{}}
   \STATE $m_{new}$, the first machine in $\textit{\textrm{MATCHLIST}}$, requests to the first element in $PL_{NH}$ (shown by $NH_{curr}$)
    \IF {$NH_{curr}$ has enough connection capacity}
        \STATE - Add $m_{new}$ to $PL_{AP}$ of $NH_{curr}$,
        \IF{$NH_{curr} \in M^i$ \AND $NH_{curr}$ does not exist in $\textit{\textrm{MATCHLIST}}$}
       	   \STATE  - Add $NH_{curr}$ to $\textit{\textrm{MATCHLIST}}$. 
        \ENDIF
    \ELSIF {  $m_{new}$ demand capacity $>$  $m_{curr}^{min}$ demand capacity ($m_{curr}^{min}$ is the current $m$ with minimum capacity)}
    	\STATE (For $m_{curr}^{min}$)
    	\STATE - Delete $m_{curr}^{min}$ from $PL_{AP}$ of $NH_{curr}$,
		\STATE - Delete $NH_{curr}$ from $PL_{NH}$ of $m_{curr}^{min}$,
		\STATE - Add $m_{curr}^{min}$ to $\textit{\textrm{MATCHLIST}}$.
		\STATE (For $m_{new}$)  \algcost{}{}
        \STATE - Add  $m_{new}$ to $PL_{AP}$ of $NH_{curr}$, 
        \STATE - Delete  $m_{new}$ from $\textit{\textrm{MATCHLIST}}$.
    \ELSIF{ $m_{new}$ demand capacity $\leq m_{curr}^{min}$ demand capacity}
    	\STATE (For $m_{new}$) 
    	\STATE - Delete $NH_{curr}$ from $PL_{NH}$ of $m_{new}$.
    \ENDIF
    \IF{$PL_{NH}$ of $m_{new} = \phi$}
       	\STATE  - Delete $m_{new}$ from $\textit{\textrm{MATCHLIST}}$, 
       	\IF{$m_{new} \in M^i$}
       		\STATE - Delete all of machines in $PL_{AP}$ of $m_{new}$. 
       	\ENDIF
     \ENDIF
    \ENDWHILE  
\end{algorithmic}
\end{algorithm}

\setcounter{algorithm}{1}
\begin{algorithm}[!htb] 
\caption{Proposed Decentralized Matching Based Relay Selection Algorithm (cont.): Step 2.}
\label{alg:DMRSA2}
\begin{algorithmic}[1]
\algsetup{linenosize=\small}
\scriptsize
        \TITLE{ \textbf{Step 2: Finish matching section and Start data forwarding}} 
        
        \STATE - Match each source with the first element in the \textit{Preference List} of next hops, 
        \STATE - Match each relay with the first element in the \textit{Preference List} of next hops, and with the first element in the \textit{Preference List} of applicant machines,
        \STATE  - Match base station with the first element in the \textit{Preference List} of applicant machines, 
        \STATE - Any machine sends data to the matched next hop.
\end{algorithmic}
\end{algorithm}

\subsubsection{\textbf{Complexity of MRSA}}\label{sec:Comp_mrsa}

%
%

MRSA can be implemented in a decentralized way for each source. We applied the deferred acceptance procedure\cite{CASMGaSh1962} to implement this algorithm (Algorithm \ref{alg:DMRSA1}). In worst case, complexity of this procedure is $O(|\textrm{MATCHLIST}|^2)$. In MRSA, $|\textrm{MATCHLIST}|=N_s+N_r$, so the complexity of MRSA is $O((N_s+N_r)^2)$. If the number of relays ($N_r$) is constant and the number of sources is equal to $N_s = n$, the complexity of MRSA is $O(n^2)$.



\subsection{Proof of stability and optimal stability of MRSA }\label{sec:opti_MRSA_proof}

This proposed decentralized algorithm is based on the deferred acceptance procedure. It is proved that the result of this algorithm is a \textbf{stable} solution \cite{CASMGaSh1962}.


The following definitions are required in the proofs:
\begin{itemize}
\item[-] Definition (in terms of matching theory): In a \textbf{stable matching}, there are no two nodes that they want each other but they match with another node.
\item[-] Definition (in terms of matching theory): The \textbf{possible matching} between an applicant node and another node means there exists at least one stable matching that assigns the applicant node to the other node.
\end{itemize}

\subsubsection{\textbf{Stable Result}}

We claim that after algorithm \ref{alg:DMRSA1} is finished, the achieved matching result will be \textbf{stable}. 


\textbf{Proof}: It is demonstrated by contradiction. We assume the proposed matching result is not stable, so there are two nodes, for example, $i$ and $j$, that prefer each other to the current matched node. Therefore, applicant node $i$, before requesting to the current matched node, has requested to node $j$ and node $j$ rejected node $i$. This means that node $j$ prefers current matched node to node $i$. Thus it is a contradiction and the provided matching is stable.

It is important to note that in order to achieve stability in this procedure, it is necessary that the device\textquotesingle s priority is not the same when selecting a path. In our scenario, according to a random location and channel condition between devices, the probability of equal capacity between two devices is near to zero. Therefore, it does not hinder the proof of the stability of the problem.




\subsubsection{\textbf{Optimal Stable Result for Sources}}


Moreover, we claim for each source (as applicant node in matching theory), the provided stable matching is at least as well as any other stable possible matching using the same nodes. 



\textbf{Proof}: We prove by induction. By the induction assumption, it is assumed that up to some point there is no applicant node that is rejected by a recipient node which was possible for the applicant node. Now, consider that at this point, an applicant node $A$ is rejected by recipient node $R$. Now, we prove for the induction step that $R$ is impossible for $A$. The recipient node keeps $q$ of the best requests ($q$ is the quota of the recipient node), such as $s_1, ..., s_q$ and other requests, such as $A$, are rejected. It is clear that for each of $s_i$ where ${1 \leq i \leq q}$, $s_i$ prefers $R$ to another recipient node except those that have rejected $s_i$. 
Now, we assume by contradiction that $R$ is possible for $A$ (i.e. a stable matching exists which has matched $A$ with $R$). It is clear that in this matching, at least one of the $s_i$s will be matched to a recipient with lower preference for it and rejected by $R$. However this matching is unstable because $s_i$ and $R$ could be matched which is preferred by both of them (due to the induction assumption). This means that the matching is unstable, and this is a contradiction. Therefore $R$ is impossible for $A$ and this proves the induction step. 
Therefore, we proved by induction that the matching is an optimal stable matching for sources.

In the following, we investigate the simulation results of the proposed algorithms in comparison with the direct transmission of data without a relay selection algorithm, as well as a completely random selection algorithm.

\section{Simulation Results}\label{sec:simulation}



To evaluate the performance of the proposed relay selection algorithms, we simulate our algorithms in a square environment, $590 \times 590$ ($m^2$). In this square, $N$ machines are randomly placed with a uniform distribution. Because of the random nature of the scenarios, the algorithm runs $n=1000$ times and the average value is provided by considering these runs. In each run, the number of sources is constantly $N^a$, and the rest of the $N^i$ machines relays, where $N^i=N-N^a$. In these simulations, the LTE uplink frequency range is considered $1900-1920 (\mathrm{MHz})$, and the WiFi uplink frequency range for outdoor connections is considered $5590-5610 (\mathrm{MHz})$.


The simulations are implemented in the C++ language. The simulations were executed on a device with a 4-core Intel(R) CPU (Intel Core i7-4710HQ @ 2.50GHz) and 16 GB of RAM. The simulation parameters are given in Table \ref{tbl:SimPar}.

We compare the proposed relay selection algorithms (ORSA and MRSA) results with two baseline algorithms namely the direct transmission Without any Relay Selection Algorithm (WRSA) and the fully Random Relay Selection Algorithm (RRSA). 

WRSA is a decentralized algorithm. In WRSA, the sources do not select any relays and only select the neighbor base station (i.e. base stations that can transmit and receive messages to the source). In this algorithm, each source requests to the neighbor base stations and creates a connection if possible. For each applicant source, if the base station has connection capacity, it accepts the source and they are assigned to each other. Hence, if the number of sources is of order $n$, the complexity of the algorithm will be $O(n)$.


RRSA is also implemented in a decentralized way. In RRSA, each source selects its next hop completely randomly only once among all the relays and the base station. In RRSA, each source only requests a single relay or the base station, and if the request was possible a connection would be established, otherwise, no other request would be made. If the source selects a relay, and the relay is unable to communicate with the base station for any reason, such as lack of connection capacity or inability to communicate with the base station, the source will not change its selected choice. Additionally, if the source selects the base station, and the base station can not communicate with the source, the source will not change its selected choice. Thus the complexity of RRSA is $O(n)$.

It is noticeable that in WRSA only one type of RF interface (LTE) is used for direct communication of the machines to the base station, But when using relays in ORSA, MRSA, and RRSA, according to the static setting of the RF interfaces, two types of interfaces (LTE and WiFi) are used simultaneously and communication between machines does not affect the direct communication of the machines to the base station.

Table \ref{tbl:SimPar} provides the default parameters in our simulations. In the following, we compare the algorithms in different aspects and investigate the impact of some different parameters by changing the default values of parameters in the different scenarios that are discussed below.


\begin{table}
   \scriptsize
   \renewcommand{\arraystretch}{1.3}
  \begin{center}
    \caption{Simulation Parameters.}
    \label{tbl:SimPar}
    \begin{tabular}{|l|c|} 
    \hline
      \textbf{Parameter} & \textbf{Default Value} \\
      \hline
      WiFi Uplink Central Frequency & $5600 \mathrm{MHz}$ \\
      WiFi Total Bandwidth & $20 \mathrm{MHz}$ \\
      Number of WiFi Channel & $1$ \\
      WiFi Power Transmission of Machines & $0.1 w$ \\
      WiFi Power Transmission of Base station & $0.1 w$ \\
      LTE Uplink Central Frequency & $1910 \mathrm{MHz}$ \\
      LTE Total Bandwidth & $20 \mathrm{MHz}$\\
      Number of LTE Channel ($No^{ch}_{\mathrm{LTE}}$) & $100$ \\
      LTE Power Transmission of Machines & $0.2 w$ \\
      LTE Power Transmission of Base station & $10 w$ \\
      Threshold of Received Power of Devices & $-121 dB$\\ 
      Mean of Normal Shadowing on Received Power & $0$ \\
      Std. Dev. of Normal Shadowing on Received Power & $4$ \\ 
      Number of Simulation Runs & $1000$ \\
      Number of Machines ($N$) & $100$ \\
      Number of Sources ($N_s$) & $0 \quad to \quad 100$ \\
      Number of Relays ($N_r$) & $N-N_s$ \\
      Length of Test Environment & $590 m$  \\
     Width of Test Environment & $590 m$ \\
      \hline
    \end{tabular}
  \end{center}
\end{table}

The different scenarios are described below:

\begin{itemize}

\item \textbf{Scenario 1}: The aim of this scenario is to compare all algorithms when the total number of network machines is constant and the number of relays and sources change proportionally.
\item \textbf{Scenario 2}: In this scenario, all algorithms are compared where the number of relays is constant and the number of sources is varied in a specific range.

\item \textbf{Scenario 3}: The impact of changing the relay number for ORSA and MRSA is investigated in this scenario.
\item \textbf{Scenario 4}: The impact of changing the number of the base station LTE channels for ORSA and MRSA is investigated in this scenario.
\end{itemize}

The evaluation metrics considered in the scenarios are as follows:
 
\begin{itemize}
\item[-] \textbf{Average Capacity of Connections between Sources and the Base station}: The capacity of each link is the maximum bit rate that can be used according to the link conditions. The average capacity of connections between sources and the base station after completing the execution of the algorithms. Hereinafter, this parameter is briefly referred to as the average capacity of sources. 

\item[-] \textbf{Average Number of Unmatched Sources}: The average of the number of sources that have not been matched after completing the execution of the algorithms.

\item[-] \textbf{Actual Execution Time of Proposed Algorithms}:  The average duration time of the proposed algorithms (ORSA and MRSA) execution is measured and recorded for all scenarios. 
The curves of time values are plotted as a scatter with smooth lines and markers.
To better visualize the trend of these curves, we calculated the trendline of these time curves using the trendline feature of Microsoft Office. We selected the polynomial trendline that had a good fit with the main curve. 

\end{itemize}

As this paper discusses uplink communication, it is important to transmit data of sources to their destinations via the base station as a bridge linking machines with the broadband infrastructure network. Therefore, the evaluation metrics include the average capacity of connections between matched sources and the base station and the number of unmatched sources.




%
%
%
%
%
%
%
%
	
	\subsection{\textbf{Scenario 1}}\label{sc:S1}



In this scenario, we investigate the difference of four algorithms, WRSA, RRSA, ORSA and MRSA, assuming the number of the base station LTE channels is a large enough constant and it does not restrict source assignment to it, directly or by one hop. Besides, the total number of network machines is constant and the number of relays and sources change proportionally. 


We compare WRSA, RRSA, ORSA and MRSA in terms of the average capacity of sources, the number of unmatched sources and the complexity in equal conditions compared as in Table \ref{tbl:s1_SimPar}.


\begin{table}[H]
   \scriptsize
   \renewcommand{\arraystretch}{1.3}
  \begin{center}
    \caption{Simulation Parameters of Scenario 1.}
    \label{tbl:s1_SimPar}
    \begin{tabular}{|l|c|c|} 
    \hline
      \textbf{Parameter} & \textbf{Value} & \textbf{Constant/Variable}\\
      \hline
      \hline
      $N$ & $100$ & Constant\\
      $N_s$ & $0 .. 100$ & Variable\\
      $N_r$ & $N-N_s$ & Variable\\
      $No^{ch}_{\mathrm{LTE}}$ & $100$ & Constant\\
      \hline
    \end{tabular}
  \end{center}
\end{table}


 \begin{enumerate}

	\item \textbf{Capacity}
			


Fig.~\ref{fig:s1Cap} shows a comparison of the average capacity of sources of WRSA, RRSA, ORSA and MRSA. The curves of ORSA, MRSA and RRSA are in a descending trend. This is because of the following reasons:


\begin{enumerate}

\item[-] increasing the number of sources along the graph increases the interference of the transmitters (sources) on the shared WiFi channel,
 and
\item[-] decreasing the number of relays across the graph reduces the improvement of capacity created due to the help of the relays.

\end{enumerate}


Due to path loss, shadowing and fading effects, attenuation of direct communication between the sources and the base station occurs. ORSA and MRSA, which select the relays to achieve optimal or stable assignment respectively, generally yield more average capacities of sources than WRSA and RRSA.

On the other hand, ORSA and MRSA converge to WRSA at the end of the curve because these algorithms have similar functionality in the absence of any relays. Furthermore, WRSA curve almost follows a constant trend along the graph. This is due to the fact that all of the assigned sources directly connect to an LTE channel of the base station with the same bandwidth. The small possible difference of WRSA points is related to the different channel conditions with each of the sources.




Additionally, as can be observed, ORSA has the best average capacity of sources among these algorithms. This is a direct result of the optimality of ORSA compared to other algorithms (that is shown in Section \ref{sec:corsa_alg}). MRSA has a result very close to the optimal result, which is at most 3\% less than ORSA result. Moreover, MRSA result is higher than WRSA and RRSA results, about 56\% and 117\% on average, respectively. Furthermore, WRSA has the next rank about this parameter results and it is superior to RRSA, about 45\% on average.

To verify our results, we analyze the results of each of the relay selection algorithms. The standard deviation of the results on all runs of the algorithms are 0.27, 0.34, 0.16 and 0.16 for WRSA, RRSA, ORSA and MRSA, respectively.






			\begin{figure}
				\centering
				\includegraphics[scale=0.5]{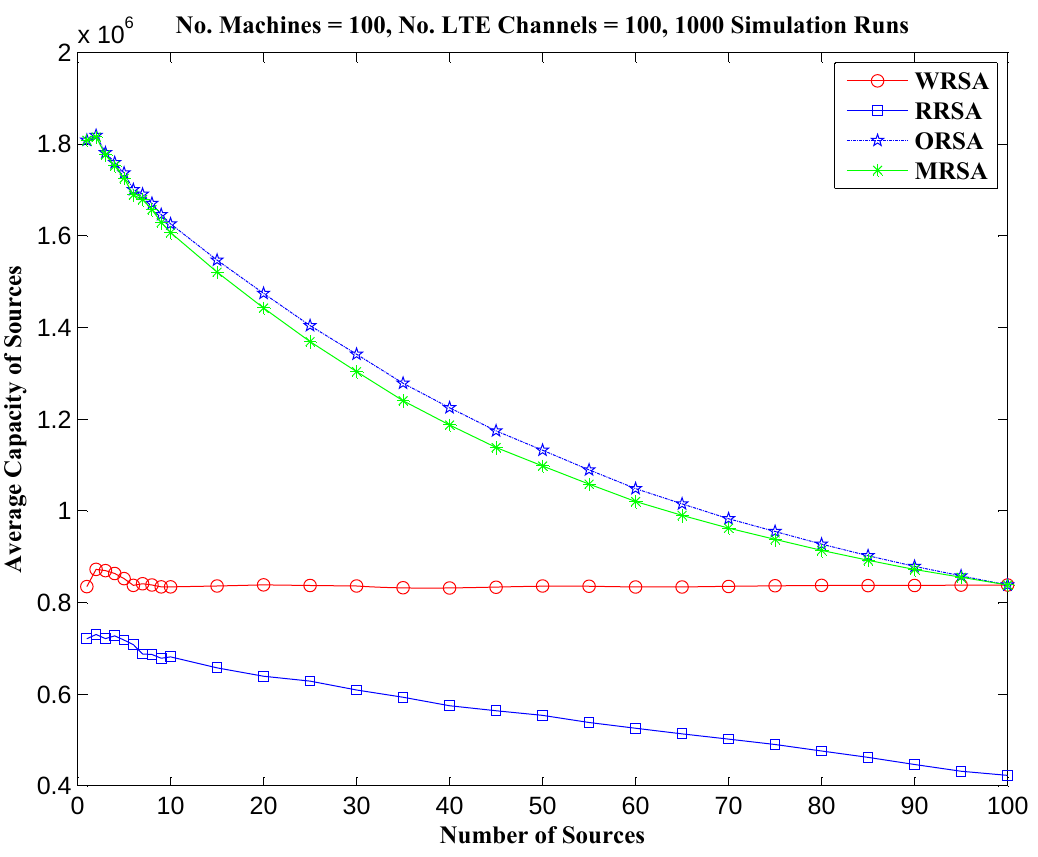}
				\caption[The average capacity of sources for WRSA, RRSA, ORSA and MRSA vs. the number of sources in Scenario 1.]{The average capacity of sources for WRSA, RRSA, ORSA and MRSA vs. the number of sources in Scenario 1.}
				\label{fig:s1Cap}
			\end{figure}

	\item \textbf{Unmatched Source Number}

%
%
%

Due to the fading and shadowing effects in all scenarios, as shown in Fig.~\ref{fig:s1Unmatched}, a number of sources still are not connected to the base station. This can be explained in more details as follows:





\begin{itemize}

\item[-] Using WRSA, all sources try to connect directly to the base station. However, as can be seen from the simulation results, the presence of fading and shadowing effects attenuates the direct connection channels. Without a relay, there is no other way to connect the sources to the base station when the direct channel conditions are poor. Therefore, as the number of sources increases, an approximately constant proportion of them are not matched to the base station.


\item[-] Using RRSA, according to the fully random nature of RRSA, each source selects between the direct connection or connection through the relays to the base station with 50-50\% probability. In other words, a request to connect any source to connect to the base station occurs in two forms:

 \begin{itemize}
  \item[-] Direct connection to the base station with a probability of 50\%, in which case the conditions will be like WRSA.
  \item[-] Connection to the base station through random selection between relays with a probability of 50\%, in which case the connection may not be established due to the lack of communication capacity of the randomly selected relay.
 \end{itemize}


Therefore, as shown in Fig.~\ref{fig:s1Unmatched}, the number of unmatched sources of the RRSA algorithm is even higher than WRSA.


%

\item[-] As mentioned in the previous section, ORSA and MRSA increase the average capacities of sources compared to WRSA and RRSA, by selecting the optimal or stable relays. As a result, it is natural that the number of unmatched sources after the execution of ORSA and MRSA is less than the other two algorithms.


Given that in Scenario 1 the total number of machines was fixed at 100, increasing the number of sources means decreasing the number of idle machines or relays. Hence up to the point where the number of sources is less than or equal to 50, the number of relays is more than or equal to the number of sources. Since each relay can help a maximum of one source to send data, after this point even if each source neighbors at least one relay, some sources still are not able to connect to relays to compensate the weakness of their direct connection with the base station. Therefore the pace of the increase in the number of unmatched sources in the second half of the ORSA and MRSA curves is greater than in the first half.




In addition, it is clear that the optimal allocation in ORSA has been able to increase the number of unmatched sources compared to MRSA by an average of 0.5 sources and at maximum of 1.6 sources. The higher average of ORSA capacity compared to MRSA, observed in Fig.~\ref{fig:s1Cap}, is also in accordance with the aforementioned fact.



\end{itemize}

			\begin{figure}
			\centering
			\includegraphics[scale=0.5]{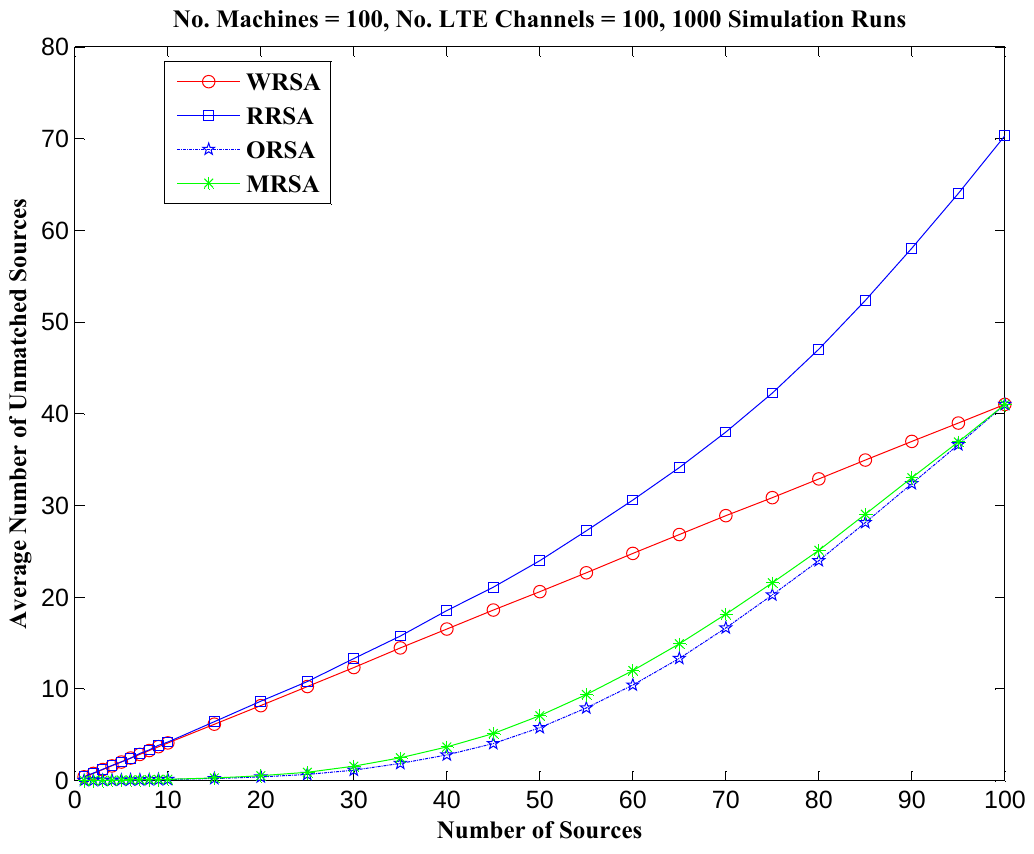}
			\caption[The average number of unmatched sources for WRSA, RRSA, ORSA and MRSA vs. the number of sources in Scenario 1.]{The average number of unmatched sources for WRSA, RRSA, ORSA and MRSA vs. the number of sources in Scenario 1.}
			\label{fig:s1Unmatched}
			\end{figure}

\item \textbf{Actual Execution Time of Proposed Algorithms}

Measured the actual execution time of our proposed algorithms is shown in Fig. \ref{fig:s1time}. In the curve trendline, the coefficient of $n^3$ is near zero and this shows that the complexity of execution time is of order $O(n^2)$. Therefore, as seen in Fig. \ref{fig:s1time}, the order of trendline of actual execution time is at least less than or equal to the complexity of both of algorithms, ORSA and MRSA, that is calculated in Subsection \ref{sec:Comp_orsa} and \ref{sec:Comp_mrsa}. Furthermore, it can be seen that ORSA actual execution time is more than MRSA actual execution time.


According to Fig.~\ref{fig:s1Cap}, the algorithms can be ordered as ORSA, MRSA, WRSA and RRSA, in terms of the average capacity, . However, in terms of complexity, WRSA and RRSA, have lower complexity than MRSA and ORSA, respectively. Therefore, these algorithms possess a trade off between the better average capacity of sources and lower complexity and vice versa.



It should be noted that in this scenario, as the number of sources increases, the number of relays decreases. As both sources and relays have an equal impact on the execution time, the worst case happens when the number of sources and the number of relays is equal in the middle point of the curve.


			
				\begin{figure}
			\centering
			\includegraphics[scale=0.47]{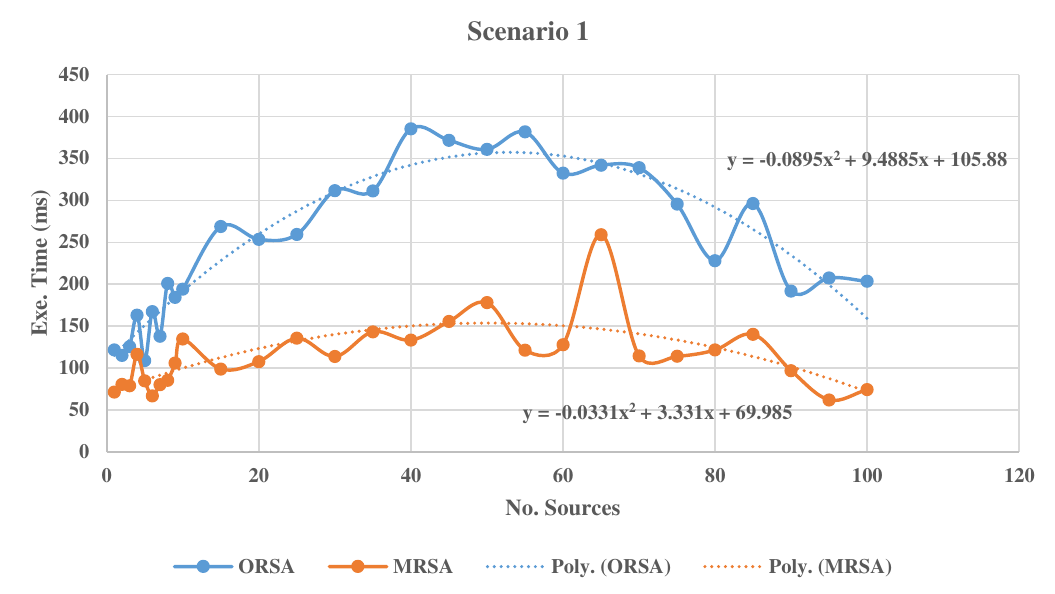} 
			\caption[The average actual execution time for ORSA and MRSA (ms) vs. the number of sources in Scenario 1.]{The average actual execution time for ORSA and MRSA (ms) vs. the number of sources in Scenario 1.}
			\label{fig:s1time}
			\end{figure} 			
			
		\end{enumerate}

			\subsection{\textbf{Scenario 2}}\label{sc:S2}




This scenario is similar to Scenario 1 with a large enough constant number of LTE channels and the number of sources is varied in a specific range. But, it has a fixed number of relays. In the following, each of the four algorithms is compared under the same conditions as in Table \ref{tbl:s2_SimPar}.


%

			
\begin{table}[H]
   \scriptsize
   \renewcommand{\arraystretch}{1.3}
  \begin{center}
    \caption{Simulation Parameters of Scenario 2.}
    \label{tbl:s2_SimPar}
    \begin{tabular}{|l|c|c|} 
    \hline
      \textbf{Parameter} & \textbf{Value} & \textbf{Constant/Variable}\\
      \hline
      \hline
      $N$ & $N_s+N_r$ & Variable\\
      $N_s$ & $0 .. 100$ & Variable\\
      $N_r$ & $75$ & Constant\\
      $No^{ch}_{\mathrm{LTE}}$ & $100$ & Constant\\
      \hline
    \end{tabular}
  \end{center}
\end{table}
			
		\begin{enumerate}

			\item \textbf{Capacity}
	


The comparison of the average capacity of sources of WRSA, RRSA, ORSA and MRSA is shown in Fig.~\ref{fig:s2Cap}. Due to the fact that WRSA does not use relays, the difference in the number of relays in Scenario 2 has no effect on its chart compared to Scenario 1. The curves of ORSA, MRSA and RRSA are in a descending trend similar to Scenario 1. Moreover, the number of relays in Scenario 2 is constant at 75, so when the number of sources is in the range $[1, 25)$, the number of relays is less than the same range in Scenario 1.  Also, in the range $[25, 100]$, the number of relays is more than the same range is in scenario 1.


Therefore, in Scenario 2, for all three algorithms in the range $[1, 25)$, there are fewer options of next hops for the sources, which makes the average capacity a less than Scenario 1. In contrast, in the range $[25, 100]$, sources have more options to select the next hop, and the average container is higher than Scenario 2.

To verify this scenario results, we analyze them by the standard deviation of the results on all runs of the algorithms. The standard deviation of the results is equal to 0.27, 0.33, 0.15 and 0.15 for WRSA, RRSA, ORSA and MRSA, respectively.

			\begin{figure}
				\centering
				\includegraphics[scale=0.5]{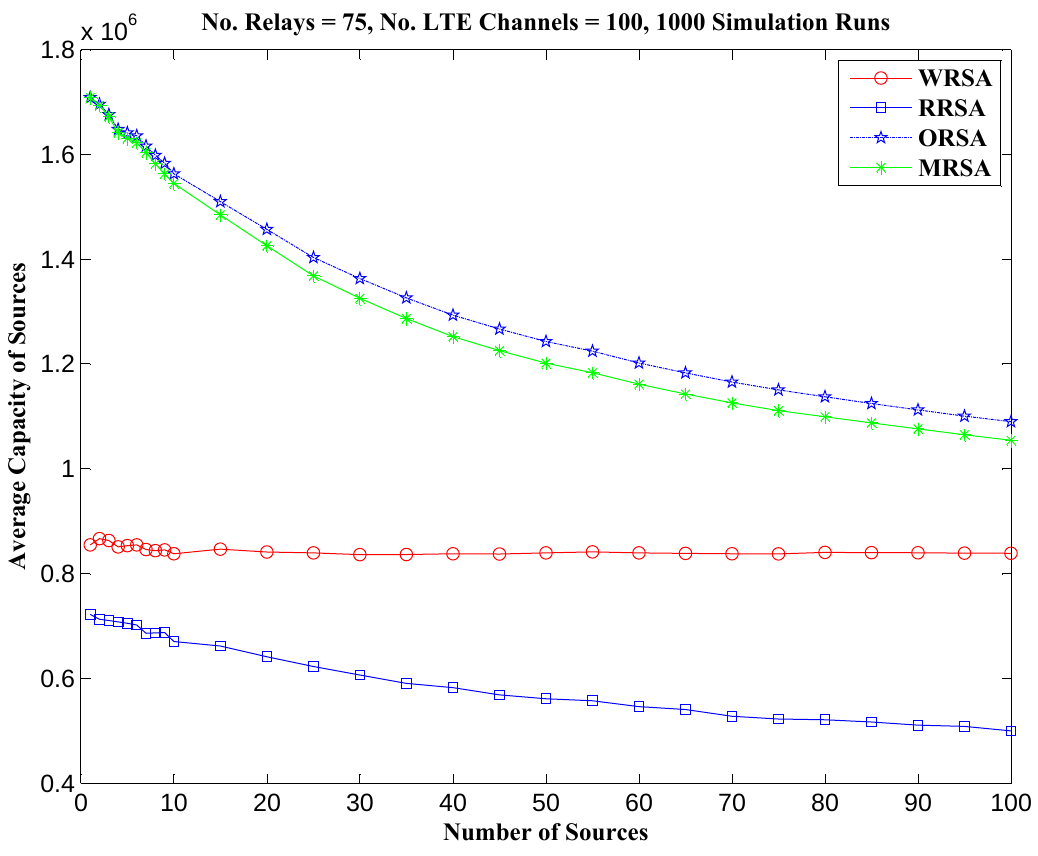}
				\caption[The average capacity of sources for WRSA, RRSA, ORSA and MRSA vs. the number of sources in Scenario 2.]{The average capacity of sources for WRSA, RRSA, ORSA and MRSA algorithms vs. the number of sources in Scenario 2.}
				\label{fig:s2Cap}
				\end{figure} 	



		\item \textbf{Unmatched Source Number}



 As mentioned in the analysis of Fig.~\ref{fig:s2Cap} , in the range $[25, 100]$, sources have more relays available to select as the next hop, and more sources can match with the base station by a relay. Therefore, ORSA, MRSA and RRSA which are dependent on relays have less unmatched sources with respect to Scenario 1. But as seen in Fig.~\ref{fig:s2unmatched}, since WRSA does not use relays, the difference in the number of relays in Scenario 2 has no effect on the number of unmatched sources compared to Scenario 1.
 



	

			\begin{figure}
			\centering
			\includegraphics[scale=0.5]{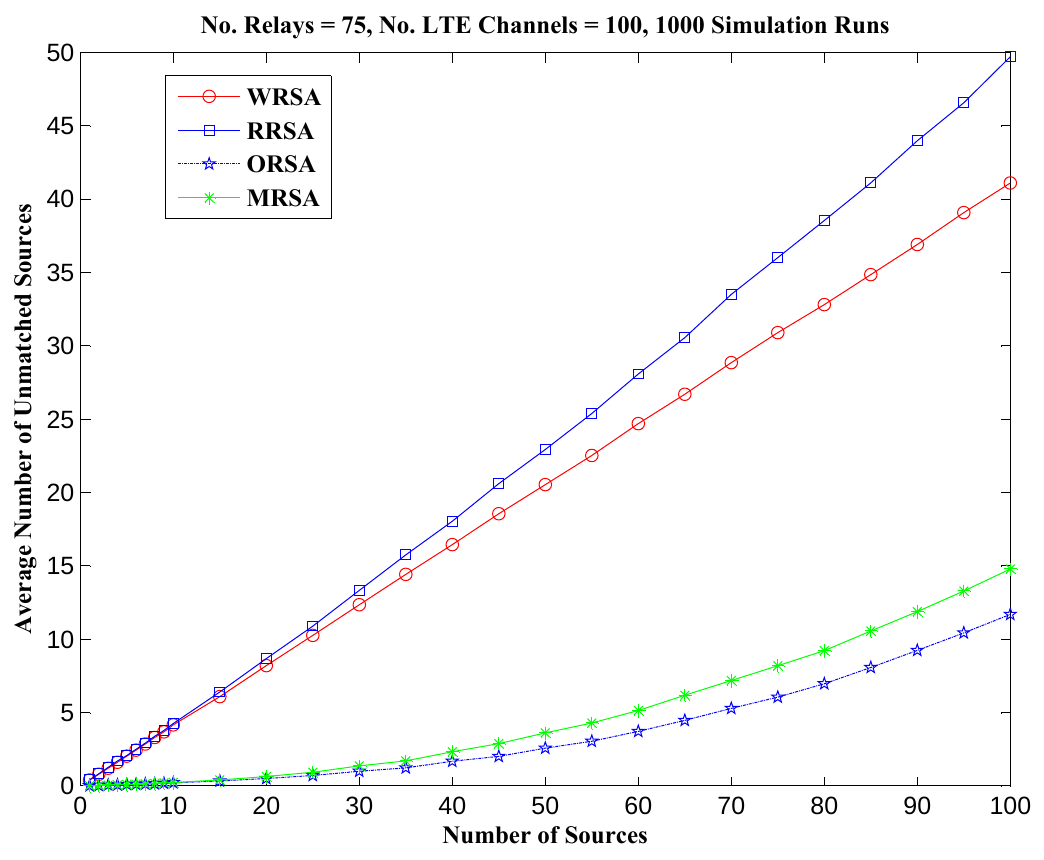}
			\caption[The average number of unmatched sources for WRSA, RRSA, ORSA and MRSA vs. the number of sources in Scenario 2.]{The average number of unmatched sources for WRSA, RRSA, ORSA and MRSA algorithms vs. the number of sources in Scenario 2.}
			\label{fig:s2unmatched}
			\end{figure} 		


\item \textbf{Actual Execution Time of Proposed Algorithms}


In this scenario, the actual execution time of our proposed algorithms is measured where the number of relays is relatively high constant ($N_r=75$). As shown in Fig. \ref{fig:s2time}, the ORSA execution time has a small coefficient of $n^2$ and the MRSA execution time is almost linear. For both of these curves, having an upper bound of $O(n^3)$ for ORSA and $O(n^2)$ for MRSA is observed, which was previously discussed in subsection \ref{sec:Comp_orsa} and \ref{sec:Comp_mrsa} respectively.



%

				\begin{figure}
			\centering
			\includegraphics[scale=0.47]{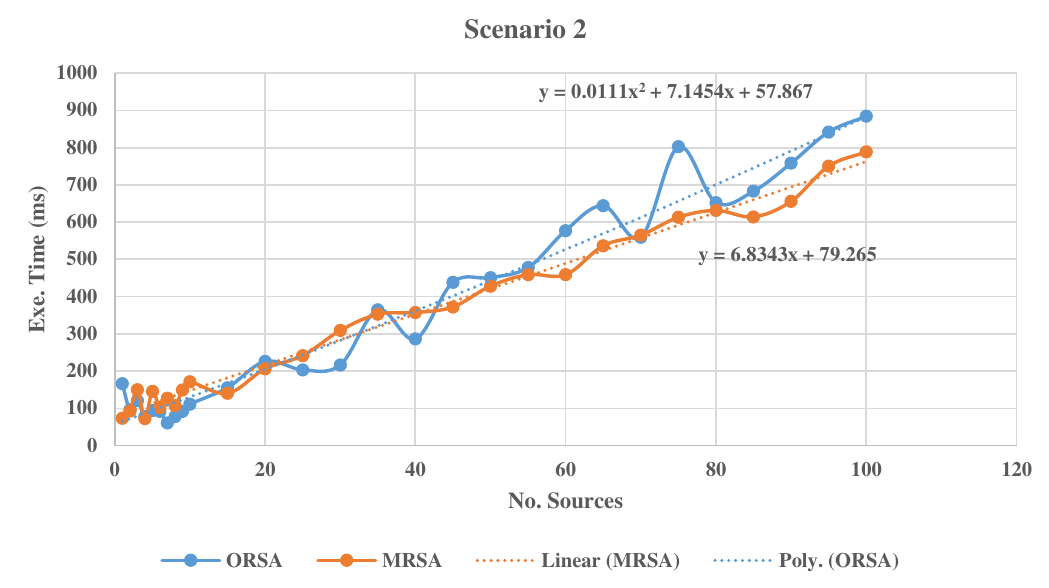}
			\caption[The average actual execution time for ORSA and MRSA (ms) vs. the number of sources in Scenario 2.]{The average actual execution time for ORSA and MRSA (ms) vs. the number of sources in Scenario 2.}
			\label{fig:s2time}
			\end{figure}

		\end{enumerate}

	\subsection{\textbf{Scenario 3}}\label{sc:S3}
		



The purpose of simulating this scenario is to investigate the effect of changing the number of relays on both ORSA and MRSA. We simulate ORSA and MRSA according to the parameters in Table \ref{tbl:s3_SimPar}.



\begin{table}[htb]
   \scriptsize
   \renewcommand{\arraystretch}{1.3}
  \begin{center}
    \caption{Simulation Parameters of Scenario 3.}
    \label{tbl:s3_SimPar}
    \begin{tabular}{|l|c|c|} 
    \hline
      \textbf{Parameter} & \textbf{Value} & \textbf{Constant/Variable}\\
      \hline
      \hline
      $N$ & $N_s+N_r$ & Variable\\
      $N_s$ & $0 .. 100$ & Variable\\
      $N_r$ & $25, 50, 75$ & Constant for Each Curve\\
      $No^{ch}_{\mathrm{LTE}}$& $100$ & Constant\\
      \hline
    \end{tabular}
  \end{center}
\end{table}
	
				\begin{enumerate}
		
					\item \textbf{Capacity}

The results shown in Fig.~\ref{fig:s3Cap} state that for ORSA and MRSA in equal conditions, a greater number of relays results in a more sources connecting to the base station. This yields in a higher average capacity of sources in the network. Additionally, similar to Scenario 1 (in Section \ref{sc:S1}), ORSA results are slightly higher than MRSA in the absence of a channel constraint. The average standard deviation of the average capacity when the number of channels are 25, 50 and 75 are equal to 0.21, 0.17 and 0.15 for ORSA and 0.21, 0.18 and 0.16 for MRSA, respectively.

			\begin{figure}
				\centering
				\includegraphics[scale=0.5]{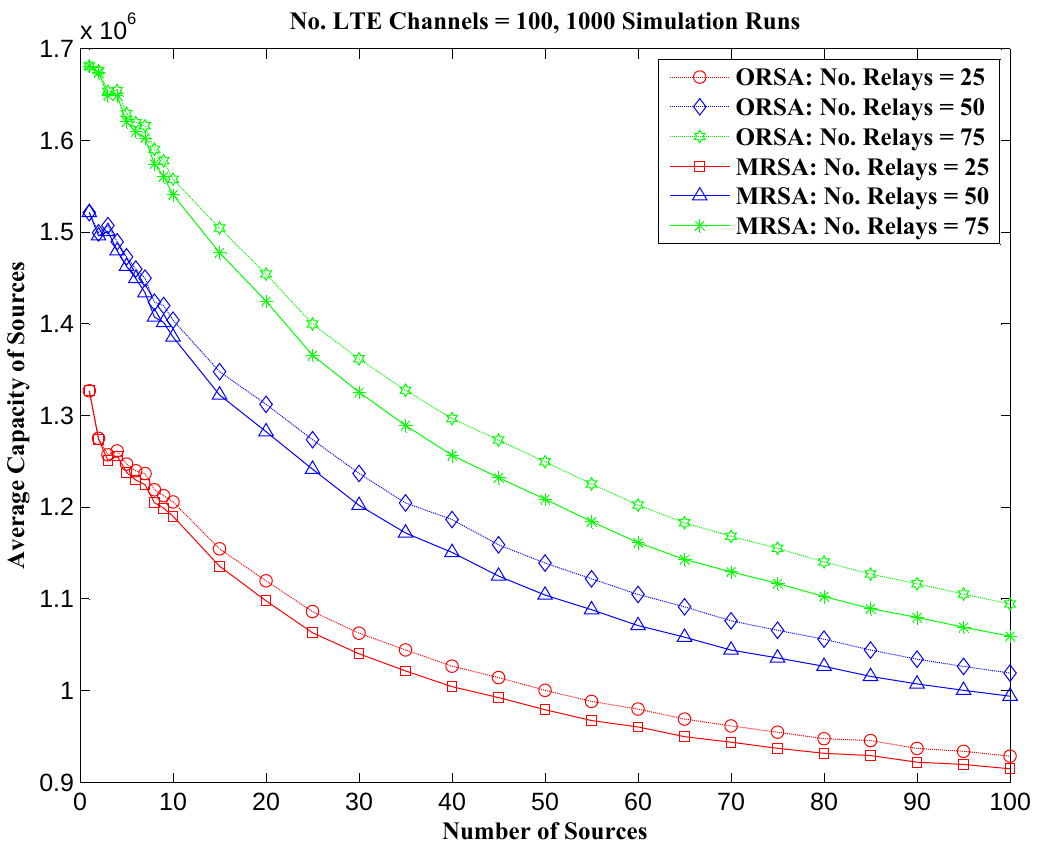}
				\caption[The average capacity of sources for ORSA and MRSA vs. the number of sources in Scenario 3.]{The average capacity of sources for ORSA and MRSA vs. the number of sources in Scenario 3.}
				\label{fig:s3Cap}
				\end{figure} 	

	
	\item \textbf{Unmatched Source Number}
	
As shown in Fig.~\ref{fig:s3unmatched}, Existence of more relays in the network will connect more sources to the base station and reduce the number of unmatched sources. As expected, ORSA unmatched sources are lower than MRSA.


			\begin{figure}
			\centering
			\includegraphics[scale=0.5]{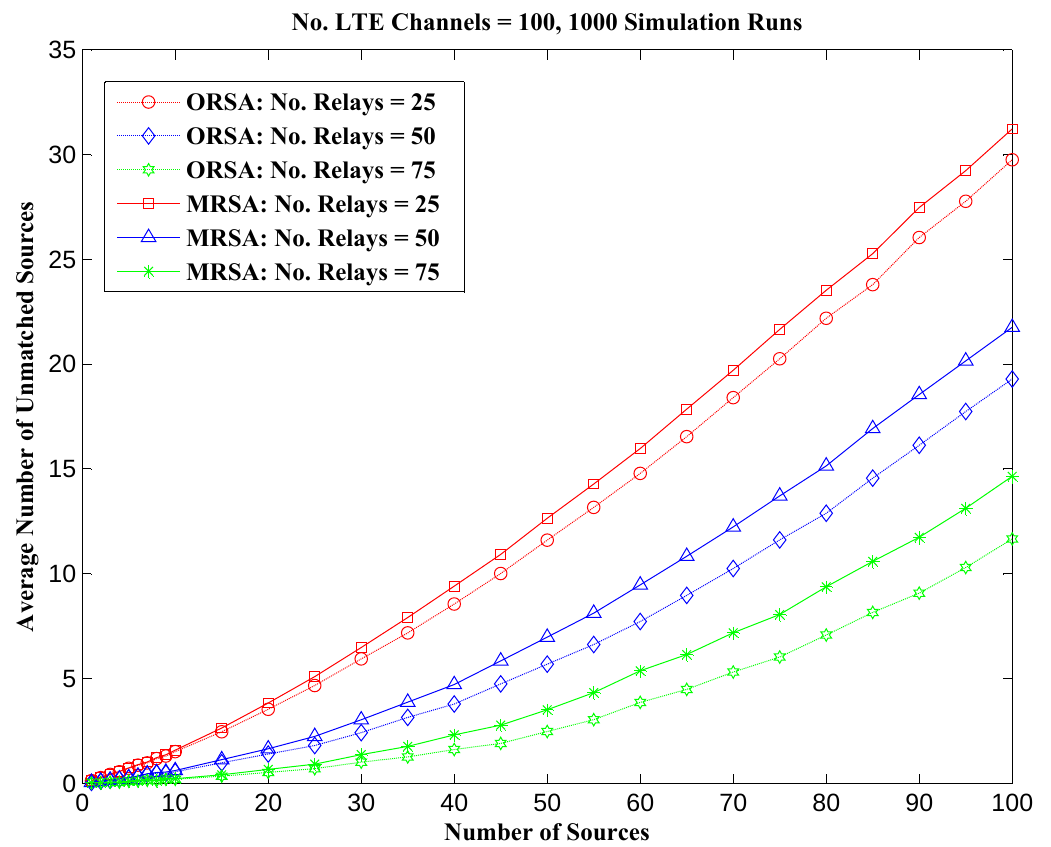}
			\caption[The average number of unmatched sources for ORSA and MRSA vs. the number of sources in Scenario 3.]{The average number of unmatched sources for ORSA and MRSA vs. the number of sources in Scenario 3.}
			\label{fig:s3unmatched}
			\end{figure} 			
		

\item \textbf{Actual Execution Time of Proposed Algorithms}

As see in Fig. \ref{fig:s3time}, the average actual execution time trendline calculated in multiple settings ($N_r=25$, $N_r=50$ and $N_r=75$) of Scenario 3 simulations indicates a small coefficient of $n^2$ for ORSA and a linear trendline for MRSA. Therefore, having an upper bound of $O(n^3)$ for ORSA and an upper bound of $O(n^2)$ for MRSA, as mentioned in subsection \ref{sec:Comp_orsa} for ORSA and subsection \ref{sec:Comp_mrsa} for MRSA is confirmed.


				\begin{figure}
			\centering
			\includegraphics[scale=0.47]{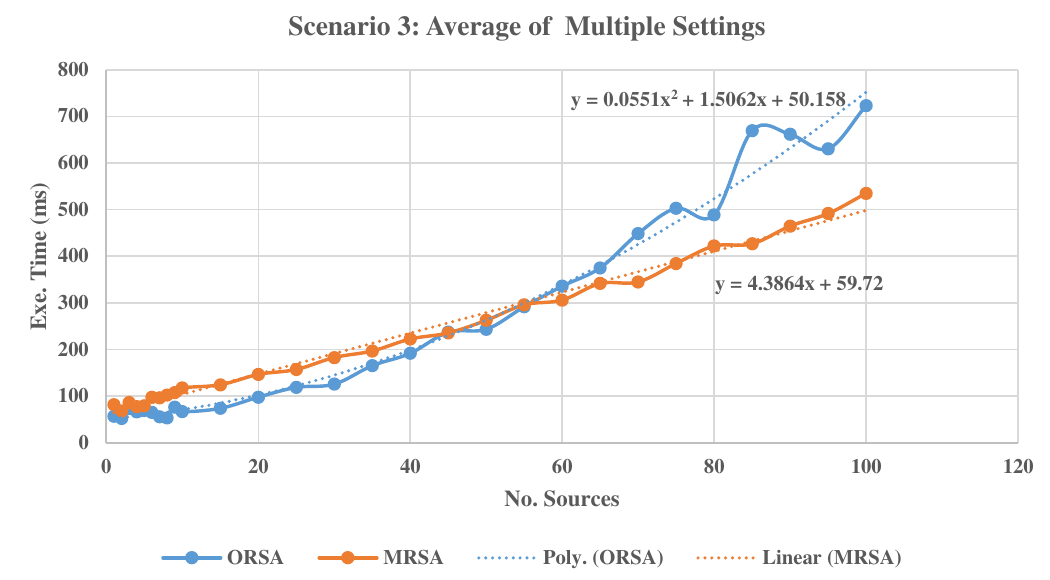}
			\caption[The average actual execution time for ORSA and MRSA in multiple settings (ms) vs. the number of sources in Scenario 3.]{The average actual execution time for ORSA and MRSA in multiple settings (ms) vs. the number of sources in Scenario 3.}
			\label{fig:s3time}
			\end{figure} 

				\end{enumerate}

	\subsection{\textbf{Scenario 4}}\label{sc:S4}



In this scenario, we want to evaluate the influence of changing the number of the base station LTE channels for ORSA and MRSA. Other parameters of this scenario are listed in Table \ref{tbl:s4_SimPar}.


\begin{table}[H]
   \scriptsize
   \renewcommand{\arraystretch}{1.3}
  \begin{center}
    \caption{Simulation Parameters of Scenario 4.}
    \label{tbl:s4_SimPar}
    \begin{tabular}{|l|c|c|} 
    \hline
      \textbf{Parameter} & \textbf{Value} & \textbf{Constant/Variable}\\
      \hline
      \hline
       $N$ & $N_s+N_r$ & Variable\\
      $N_s$ & $0 .. 100$ & Variable\\
      $N_r$ & $100$ & Constant\\
      $No^{ch}_{\mathrm{LTE}}$ & $25, 50, 75$ & Constant for Each Curve\\
      \hline
    \end{tabular}
  \end{center}
\end{table}

				\begin{enumerate}
		
						\item \textbf{Capacity}

The results of the simulation of this scenario are presented in Fig.~\ref{fig:s4Cap}. These curves illustrate the following observations:


\begin{itemize}

\item[-] Regarding the fact that the total LTE bandwidth has a constant value ($20 \mathrm{MHz}$ as stated in Table \ref{tbl:SimPar}), a lower number of channels causes a greater capacity portion for each source. The optimal results show that despite a lower channel number in total, the ORSA-25 channel curve has the highest average capacity.


\item[-] Optimal results provided by ORSA, show that when the number of LTE channels is 25 the average capacity of sources is higher than when the number of LTE channels is 50, and when the number of LTE channels is 50 the average capacity of sources is higher than when the number of LTE channels is 75.


\item[-] When the number of sources reaches the number of channels (25, 50 and 75), all curves show the average capacity drop. This is due to the fact that when the number of sources exceeds the number of available LTE channels some sources to be unable to connect to the base station.

\end{itemize}

The average standard deviation of the average capacity when the number of channels are 25, 50 and 75 are equal to 0.14, 0.13 and 0.14 for ORSA and 0.14 for MRSA in all cases., respectively.



%

			\begin{figure}
				\centering
				\includegraphics[scale=0.5]{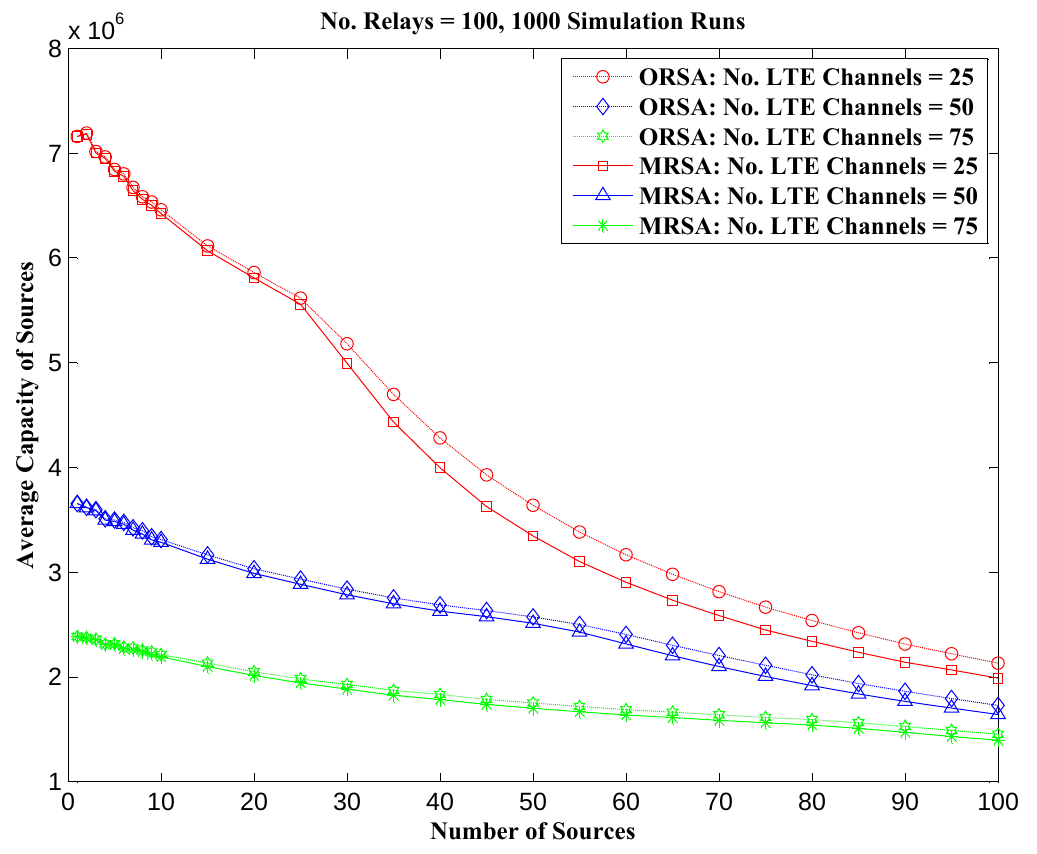}
				\caption[The average capacity of sources for ORSA and MRSA vs. the number of sources in Scenario 4.]{The average capacity of sources for ORSA and MRSA vs. the number of sources in Scenario 4.}
				\label{fig:s4Cap}
				\end{figure} 		


			\item \textbf{Unmatched Source Number}


As it can be seen in Fig.~\ref{fig:s4unmatched}, when the base station has fewer LTE channels to communicate with machines, the number of sources able to communicate with the base station decreases and the number of unmatched sources increases.


Also, The simulation results show that in both ORSA and MRSA, as long as the number of sources is less than the number of LTE channels available, most sources can be connected to the base station, and only a small number remain unaddressed due to conditions such as channel attenuation. Afterwards, bypassing the number of sources through the number of channels, it is seen that as the number of sources increases, the number of unmatched sources also increases linearly.

%
%


 the number of channels available to the sources determines the number of sources that can be connected to the base station, and the remaining sources over the number of available LTE channels are not matched.
		
For example, when the number of LTE channels is equal to 25, until the number of sources in the graph is less than or equal to 25, there are no unmatched sources. But when the number of channels exceeded the number of channels available, additional sources can not be matched.



Therefore, regardless of the type of relay selection algorithm, since the number of LTE channels of the base station determines the number of machines that can be connected to the base station, the restriction of the number of LTE channels directly affects the number of unmatched sources.

			\begin{figure}
			\centering
			\includegraphics[scale=0.5]{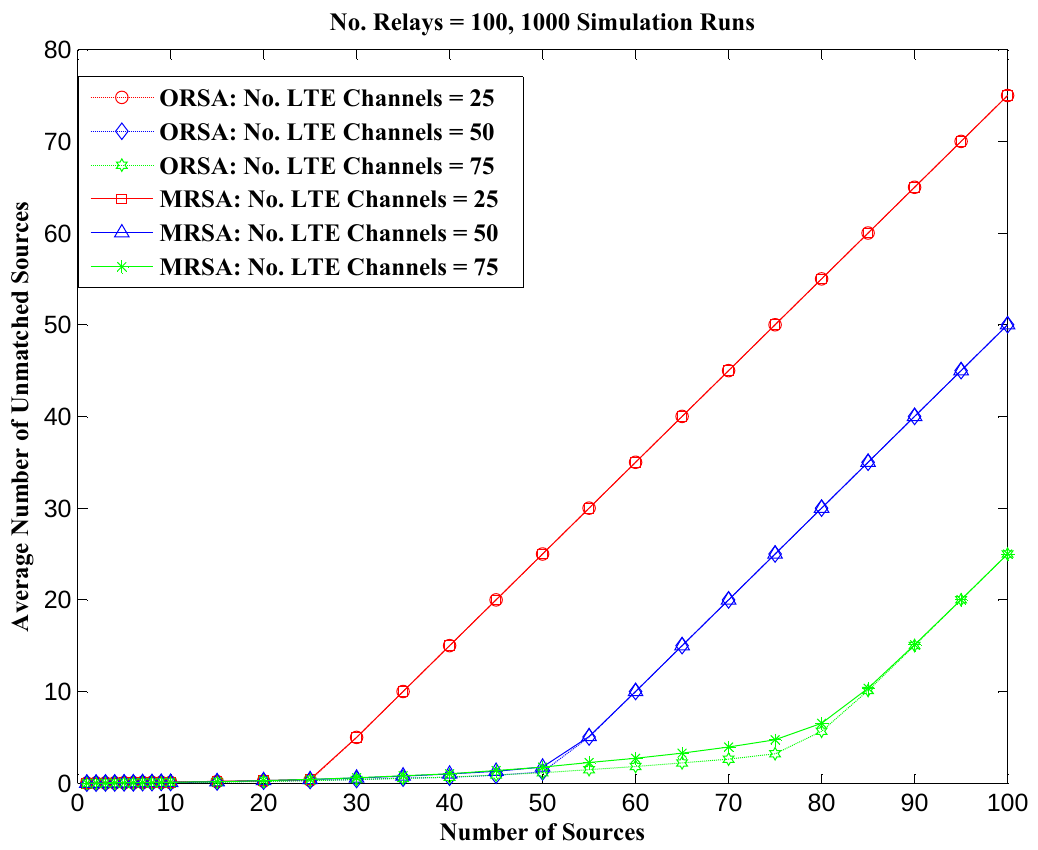}
			\caption[The average number of unmatched sources for ORSA and MRSA vs. the number of sources in Scenario 4.]{The average number of unmatched sources for ORSA and MRSA vs. the number of sources in Scenario 4.}
			\label{fig:s4unmatched}
			\end{figure} 	
		


	\item \textbf{Actual Execution Time of Proposed Algorithms}

The actual execution time of ORSA and MRSA and their trendlines is shown in Fig. \ref{fig:s4time}. In this scenario with a constant relay number ($N_r=100$) and variable source number, it is again observed that the coefficient of $n^2$ for ORSA is relatively small and the trendline of MRSA is linear. Therefore, in this scenario, as in the previous scenarios, the time complexity calculated in subsections \ref{sec:Comp_orsa} for ORSA and \ref{sec:Comp_mrsa} for MRSA is not violated.
	

				\begin{figure}
			\centering
			\includegraphics[scale=0.47]{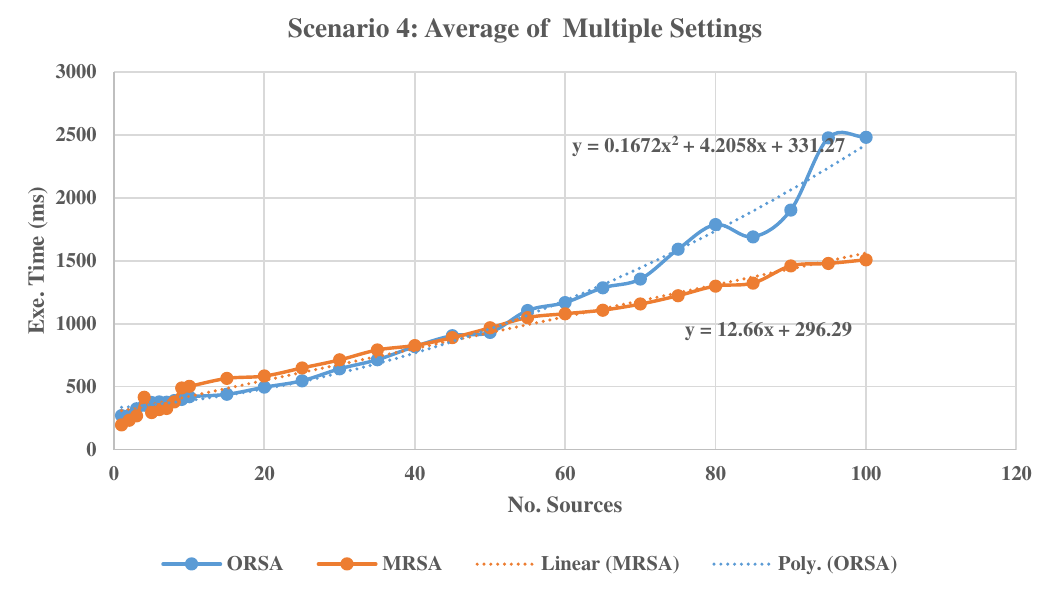}
			\caption[The average actual execution time for ORSA and MRSA in multiple settings (ms) vs. the number of sources in Scenario 4.]{The average actual execution time for ORSA and MRSA in multiple settings (ms) vs. the number of sources in Scenario 4.}
			\label{fig:s4time}
			\end{figure}

			\end{enumerate}

\section{Conclusion}\label{sec:conclusion}


In this paper, two novel algorithms were proposed for relay selection in M2M communications. The first method (ORSA) is a centralized algorithm to find the optimal relay selection. ORSA is implemented by two transformations. Throughout this algorithm a new solution for the $k$-cardinality assignment problem is provided. The second method (MRSA) is a decentralized algorithm designed based on concepts from matching theory. The result of MRSA is a stable solution for relay selection. In all of the algorithms, static RF interfaces setting is considered to allow the parallel use of interfaces for data transmission. This type of simultaneous usage can help to improve the performance of the network. In the future, a dynamic RF interfaces setting can be considered in the design of the relay selection method. The results show that ORSA has the optimal average capacity, providing solutions about 3\% higher than MRSA, when there is no restriction on the number of channels. Following ORSA, MRSA leads to solutions with average capacity higher than WRSA and RRSA, about 56\% and 117\%, respectively. Moreover, the comparison of both proposed algorithms with WRSA and RRSA, shows that ORSA and MRSA are more successful in increasing the average capacity of connections between sources and the base station and decreasing the number of unmatched sources.

\appendices


%

%

\section{Proof of stability and optimal stability of MRSA }\label{sec:opti_MRSA_proof}

This proposed decentralized algorithm is based on the deferred acceptance procedure. It is proved that the result of this algorithm is a \textbf{stable} solution \cite{CASMGaSh1962}.


The following definitions are required in the proofs:
\begin{itemize}
\item[-] Definition (in terms of matching theory): In a \textbf{stable matching}, there are no two nodes that they want each other but they match with another node.
\item[-] Definition (in terms of matching theory): The \textbf{possible matching} between an applicant node and another node means there exists at least one stable matching that assigns the applicant node to the other node.
\end{itemize}

\subsubsection{\textbf{Stable Result}}

We claim that after algorithm \ref{alg:DMRSA1} is finished, the achieved matching result will be \textbf{stable}. 


\textbf{Proof}: It is demonstrated by contradiction. We assume the proposed matching result is not stable, so there are two nodes, for example, $i$ and $j$, that prefer each other to the current matched node. Therefore, applicant node $i$, before requesting to the current matched node, has requested to node $j$ and node $j$ rejected node $i$. This means that node $j$ prefers current matched node to node $i$. Thus it is a contradiction and the provided matching is stable.

It is important to note that in order to achieve stability in this procedure, it is necessary that the device\textquotesingle s priority is not the same when selecting a path. In our scenario, according to a random location and channel condition between devices, the probability of equal capacity between two devices is near to zero. Therefore, it does not hinder the proof of the stability of the problem.




\subsubsection{\textbf{Optimal Stable Result for Sources}}


Moreover, we claim for each source (as applicant node in matching theory), the provided stable matching is at least as well as any other stable possible matching using the same nodes. 



\textbf{Proof}: We prove by induction. By the induction assumption, it is assumed that up to some point there is no applicant node that is rejected by a recipient node which was possible for the applicant node. Now, consider that at this point, an applicant node $A$ is rejected by recipient node $R$. Now, we prove for the induction step that $R$ is impossible for $A$. The recipient node keeps $q$ of the best requests ($q$ is the quota of the recipient node), such as $s_1, ..., s_q$ and other requests, such as $A$, are rejected. It is clear that for each of $s_i$ where ${1 \leq i \leq q}$, $s_i$ prefers $R$ to another recipient node except those that have rejected $s_i$. 
Now, we assume by contradiction that $R$ is possible for $A$ (i.e. a stable matching exists which has matched $A$ with $R$). It is clear that in this matching, at least one of the $s_i$s will be matched to a recipient with lower preference for it and rejected by $R$. However this matching is unstable because $s_i$ and $R$ could be matched which is preferred by both of them (due to the induction assumption). This means that the matching is unstable, and this is a contradiction. Therefore $R$ is impossible for $A$ and this proves the induction step. 
Therefore, we proved by induction that the matching is an optimal stable matching for sources.

  \section*{Acknowledgment}
The authors would like to thank Dr. Hamid Zarrabi Zade, Dr. Hossein Ajorloo, and Mr. Ahmad Haj Reza Jafar Abadi for their helpful comments and guidance.

\begin{IEEEbiography}[{\includegraphics[width=1in,height=1.25in,clip,keepaspectratio]{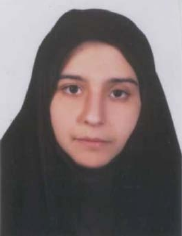}}]{Monireh Allah Gholi Ghasri}
received B.Sc. degree from Amirkabir University of Technology, Tehran, Iran, in 2012 and M.Sc. degree from Sharif University of Technology, Tehran, Iran, in 2014, both in Information Technology (IT) Engineering. Currently, she is working toward the Ph.D. degree in Computer Engineering in Wireless Networking Lab, Sharif University of Technology, Tehran, Iran. Her research interests include wireless networks, Machine-to-Machine (M2M) communications, Internet of Things (IoT), next-generation network management, relay selection, network protocols and architectures, game theory, and matching theory.
\end{IEEEbiography}


\begin{IEEEbiography}[{\includegraphics[width=1in,height=1.25in,clip,keepaspectratio]{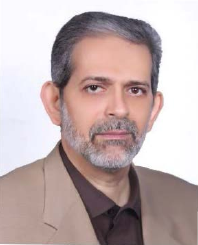}}]{Ali Mohammad Afshin Hemmatyar}
received B.Sc., M.Sc. and Ph.D. degrees in electrical engineering from Sharif University of Technology, Tehran, Iran in 1988, 1991, and 2007, respectively. Since 1991, he has joined Department of Computer Engineering in Sharif University of Technology, where he is currently an assistant professor. His research interests are Vehicular Ad-hoc Networks, Mobile Ad-hoc Networks, Cognitive Radio Networks, Wireless Sensor Networks and Internet of Things (IoT).
\end{IEEEbiography}


\begin{IEEEbiography}[{\includegraphics[width=1in,height=1.25in,clip,keepaspectratio]{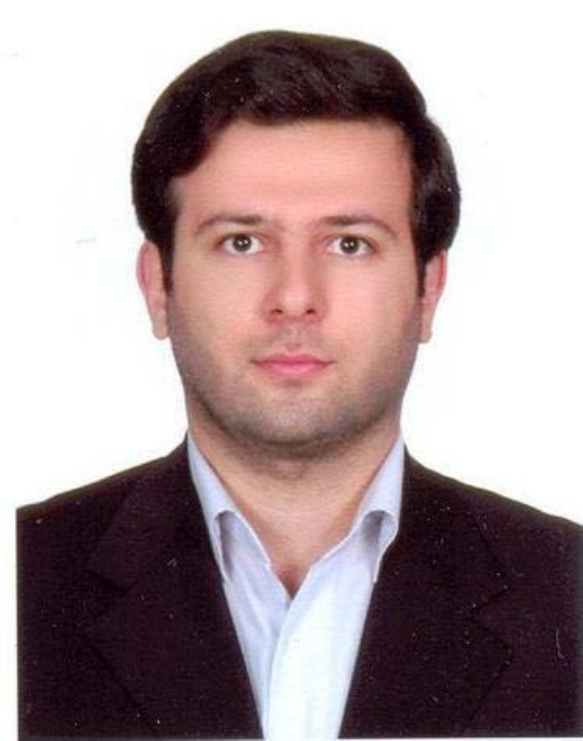}}]{Siavash Bayat}
received B.Sc. and M.Sc. degrees in electrical engineering from Sharif University of Technology, Tehran, Iran in 2002 and 2005, respectively, and Ph.D. degree in electrical engineering from University of Sydney, Australia, in 2013. He was then a Postdoctoral Research Fellow at University of Sydney, prior to joining Sharif University of Technology in 2014, where he is currently an Assistant Professor. From 2007 to 2009, he was affiliated with Sharif University of Technology, where he held a position of faculty member with responsibility for the research in wireless communication networks system design. His research interests include wireless resource management, wireless communications, Internet of Things (IoT), signal processing, heterogeneous networks, cognitive radio, game theory, and physical layer security.
\end{IEEEbiography}


\begin{IEEEbiography}[{\includegraphics[width=1in,height=1.25in,clip,keepaspectratio]{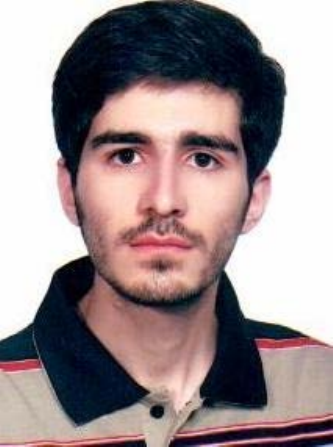}}]{Mostafa Mahdieh}
received B.Sc. and M.Sc. degrees in computer engineering from Sharif University of Technology, Iran, in 2010 and 2012, respectively. Currently, he is working toward the Ph.D. degree in Computer Engineering in Software Quality Research Lab, Sharif University of Technology, Tehran, Iran. His research interests include software engineering, software testing, machine learning, and algorithms.
\end{IEEEbiography}

\EOD

\end{document}